\documentclass[a4paper,11pt]{article}

\usepackage[T1]{fontenc}               
\usepackage[italian, english]{babel}     
\usepackage{lmodern}                   
\usepackage{textcomp}                  
\usepackage{microtype}                 

\usepackage{mathtools, amssymb, amsthm}
\usepackage{mathrsfs}                  
\usepackage{dsfont}                    
\usepackage{accents}                   
\usepackage{aliascnt}                  

\usepackage{xcolor}                    
\definecolor{myurlcolor}{rgb}{0,0,0.4}
\definecolor{mycitecolor}{rgb}{0,0.5,0}
\definecolor{myrefcolor}{rgb}{0.5,0,0}

\usepackage{graphicx}                  
\usepackage{tikz}                      
\usepackage{tikz-cd}                   
\usetikzlibrary{graphs, decorations.pathmorphing, decorations.markings}
\usepackage{caption}                   

\usepackage[pagebackref=true, colorlinks=true]{hyperref} 
\hypersetup{
    linkcolor=myrefcolor,
    citecolor=mycitecolor,
    urlcolor=myurlcolor,
}

\usepackage[capitalize, nameinlink]{cleveref}   

\usepackage{changepage}                
\usepackage{enumitem}                  
\usepackage{braket}                    

\newtheorem{theorem}{Theorem}[section]

\newcommand\nuovothm[3]{
  \newaliascnt{#1}{theorem}
  \newtheorem{#1}[#1]{#2}
  \aliascntresetthe{#1}
  \crefname{#1}{#2}{#3}
}

\nuovothm{lemma}{Lemma}{Lemmas}
\nuovothm{proposition}{Proposition}{Propositions}
\nuovothm{corollary}{Corollary}{Corollaries}
\nuovothm{definition}{Definition}{Definitions}
\nuovothm{remark}{Remark}{Remarks}
\nuovothm{assumption}{Assumption}{Assumptions}
\nuovothm{example}{Example}{Examples}
\nuovothm{question}{Question}{Questions}

\newcommand{\be}{\begin{equation}}
\newcommand{\ee}{\end{equation}}

\newcommand{\dd}{{\rm d}}
\newcommand{\de}{\partial}

\newcommand{\m}{\mathscr{M}}

\usepackage{todonotes}                 

\title{A groupoidal description of elementary particles}

\author{A. Ibort$^{1,2,6}\footnote{Corresponding author.} $ \href{https://orcid.org/0000-0002-0580-5858}{\includegraphics[scale=0.5]{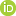}},
G. Marmo$^{3,4,7}$ \href{https://orcid.org/0000-0003-2662-2193}{\includegraphics[scale=0.5]{ORCID.png}}
A. Mas$^{2,5,8}$ \href{https://orcid.org/0000-0003-0532-0938}{\includegraphics[scale=0.5]{ORCID.png}},
L. Schiavone$^{6,9}$ \href{https://orcid.org/0000-0002-1817-5752}{\includegraphics[scale=0.7]{ORCID.png}}, \\ 
\footnotesize{$^{1}$\textit{Departamento de Matematicas, Universidad Carlos III de Madrid,}} \\ 
\footnotesize{\textit{Avenida de la Universidad 30, 28911, Legan\'es, Madrid, Spain}} \\
\footnotesize{$^{2}$\textit{Instituto de Ciencias Matem\'{a}ticas  CSIC-UAM-UC3M-UCM (ICMAT),}} \\
\footnotesize{\textit{Calle Nicol\'as Cabrera, 13--15
Campus de Cantoblanco, UAM, 28049 Madrid, Spain}} \\
\footnotesize{$^{3}$\textit{Dipartimento di Fisica, Universit\`a degli Studi di Napoli Federico II,}} \\
\footnotesize{\textit{Via Cintia, Monte S. Angelo I-80126, Napoli, Italy}} \\
\footnotesize{$^{4}$\textit{INFN Sezione di Napoli,}} \\
\footnotesize{\textit{Via Cintia, Monte S. Angelo I-80126, Napoli, Italy}} \\
\footnotesize{$^{5}$\textit{Dipartimento di Matematica e Applicazioni, Universit\`a degli Studi di Napoli Federico II,}} \\
\footnotesize{\textit{Via Cintia, Monte S. Angelo I-80126, Napoli, Italy}}  \\
\footnotesize{$^{6}$\textit{ e-mail: \texttt{albertoi@math.uc3m.es}}} \\
\footnotesize{$^{7}$\textit{ e-mail: \texttt{marmo@na.infn.it}}} \\
\footnotesize{$^{8}$\textit{ e-mail: \texttt{arnau.mas@icmat.es}}} \\
\footnotesize{$^{9}$\textit{ e-mail: \texttt{luca.schiavone@unina.it}}} 
}
\date{}
 
\begin{document}

\maketitle 


\begin{abstract}
In this work, we show that extending the standard description of space-time symmetries from groups of isometries to the more flexible framework of kinematical groupoids allows for the extension of Wigner's program to curved space-times. We propose a new definition of elementary particles as irreducible projective representations of the kinematical groupoids supporting the theory. By choosing a natural kinematical groupoid associated with any space-time, called the \textit{Wigner groupoid}, we demonstrate that such irreducible projective representations are characterized by quantum numbers similar to those characterizing the irreducible projective representations of the Poincar\'e group.

Describing the irreducible projective representations of groupoids poses its own difficulties. To address this, we develop a suitable extension of Mackey's theory of induced representations of groups, proving that projective representations of transitive Lie groupoids with connected isotropy groups are in one-to-one correspondence with the projective representations of their isotropy groups. The application of these results provides a classification of elementary particles valid for a large class of space-times. This classification largely reproduces Wigner's standard classification on Minkowski space-time, while a new family of representations emerges, corresponding to massless particles in the presence of a magnetic-like background field.
\end{abstract}

\tableofcontents


\section{Introduction: Wigner's program and the notion of elementary particle}\label{sec:introduction}

In the seminal paper \cite{Wigner-UnitaryRepresentations-1939}, \textit{ E. Wigner} identified elementary particles with irreducible projective representations of the Poincar\'e group. The Poincar\'e group $\mathcal{P}$, as the group of isometries of Minkowski space-time $\mathbb{M}^{1,3}$, determines the foundational blocks upon which a quantum theory on Minkowski space-time is built. This principle, also known as Wigner's program, provided the first classification of elementary particles by the quantum numbers $(m,s)$, representing, respectively, the mass and the spin of the particle (see Sect. \ref{sec:wigner_classification} for a succinct review of Wigner's classification). Today, it is part of the basic notions in any textbook on the theory of quantum fields (see, for instance, \cite{Weinberg-QFT1-1995, Weinberg-QFT2-1996}).

Unfortunately, such a program depends critically on the fact that the space-time used to construct the theory is Minkowski space-time or, at least, a space-time possessing a large enough group of isometries, such as (A)dS spaces. Indeed, a generic space-time\footnote{We will assume that a space-time $\m$ is a smooth connected manifold, more precisely, a smooth, Hausdorff, second countable, hence paracompact, manifold, of dimension $m = 1 + d$ carrying a semi-Riemannian metric $\eta$ of signature $(1, d)$, $+-\ldots-$, which is orientable, time orientable and strongly causal (see, for instance, \cite{Low-NullGeodesics-1989, Bautista-Ibort-Causality-2015} for details on the topology and geometrical structure of space-times).} $\mathscr{M}$ has a trivial group of isometries consisting only of the identity map. In fact, generic perturbations of the flat Minkowski metric $\eta_0$ with signature $+---$ have a trivial group of isometries. However, the physical characteristics describing elementary particles seem to be remarkably stable under such perturbations, as the accumulated experimental evidence collected over the last one hundred years confirms.

One way to tackle this difficulty is to consider the Poincar\'e group only as a local approximate symmetry. In fact, the metric $\eta$ of the given space-time $\mathscr{M}$ can be approximated locally by the Minkowski metric (for instance, by using geodesic coordinates). However, the difficulty raised before remains because, even locally, a generic space-time has a trivial group of isometries. Any space-time can be approximated locally by Minkowski space-time, but it is not isometric to it, not even locally. That is, there is no system of coordinates such that the metric $\eta$ becomes the flat Minkowski metric. Even in this scenario, we could insist that the Poincar\'e group is an ``approximate'' symmetry, but then we should clarify what it means to be an approximate symmetry of the metric $\eta$. In such a case, how does this affect the quantum numbers $(m,s)$? That is, should we conclude that the correct quantum numbers $(m,s)$ describing an actual particle on the curved space-time $\mathcal{M}$ are only well defined locally and will be of the form $m = m_0 + \delta m$, $s = s_0 + \delta s$, with $\delta m$, $\delta s$ depending on the ``difference'' between $\eta$ and $\eta_0$ on the given neighborhood we are considering?

Instead of pursuing this path, we will address the problem from a different perspective. Groups appear in physics as the mathematical way to describe symmetries, following H. Weyl's principle that symmetries are implemented via groups of transformations. However, as pointed out recently (see for instance \cite{Weinstein-GroupoidsSymmetry-1996}), groups are not always the best way to describe them. For instance, we find ourselves in this situation when studying a system that ideally would possess some global symmetry described by a group, but because of the restrictions imposed by the problem, there is no such group action, think for instance when we have to consider only a subset of the total space where the group would be acting.   When the notion of symmetry described by groups of transformations fades out, we should replace it with its natural extension, which is much more flexible and particularly well suited for this task: groupoids (see Sect. \ref{Sec: Relativistic groupoids: The Wigner groupoid} and the corresponding subsections for a detailed description of these ideas). We will follow this new conceptual approach and consider that the symmetries of a given space-time are not given by its group of isometries but by a groupoid associated with it. We will see that in the particular instance of Minkowski space-time both notions coincide, but in the case of curved space-times, the groupoid of symmetries of the given space-time still exists and is non-trivial, while the group of isometries becomes trivial. It will be discussed why the appropriate groupoid of symmetries of a given space-time is the so-called Wigner groupoid, carefully described in Sect. \ref{sec:wigner}.

Following this idea, we will propose a new classification of elementary particles based on the notion of irreducible projective representations of the groupoid of symmetries, the Wigner groupoid, of the given space-time (see Sect. \ref{Sec: Mackey's imprimitivity theorem and the theory of unitary representation of groupoids} for the discussion of the theory of projective representations of groupoids).

In Sect. \ref{Sec: Relativistic particles and relativistic groupoids: Wigner's elementary particles revisited}, the new proposal for the classification of elementary particles is implemented by computing explicitly the irreducible projective representations of the Wigner groupoid of an arbitrary space-time. It is found that the classification provided by such representations is remarkably similar to the classification provided by the standard representations of the Poincar\'e group, except for massless particles where significant differences are found. Before discussing this point, we would like to comment on the reason behind this similarity.

The main theorem, Thm. \ref{Thm: Mackey theorem transitive groupoids}, obtained in Sect. \ref{sec:projective_rep_groupoids}, states that the projective representations of transitive Lie groupoids are in one-to-one correspondence with the projective representations of their isotropy groups. As will be discussed in Sect. \ref{Sec: Relativistic particles and relativistic groupoids: Wigner's elementary particles revisited}, the isotropy groups of the Wigner groupoid and the ``little groups'' used in the construction of the representations of the Poincar\'e group are very similar, which explains the similarities. This fact will also explain the robustness of the notion of elementary particles under perturbations. Indeed, the classification of elementary particles obtained by the methods established in this research is valid in any space-time, flat or not, whether it is a homogeneous space or not.

Elementary particles also constitute a fundamental ingredient in relativistic quantum field theories; however, it is not the purpose of this paper to analyze how the proposed change in the notion of symmetry will affect the role of elementary particles in quantum field theories on curved space-times. This problem will be addressed in subsequent works. We must point out, though, that difficulties similar to, or worse than, the ones discussed before are present when trying to come up with a proper description of elementary particles on generic space-times. Relativistic quantum field theories are, by definition, covariant with respect to the action of the Poincar\'e group, and it is assumed that the Poincar\'e group is represented unitarily on the Hilbert space supporting the theory and leaving the vacuum state of the theory invariant (for instance, these are explicit axioms in the Streater-Wightman axiomatic approach to QFT \cite{streater2000pct}). Obviously, all these requirements break down in the absence of the Poincar\'e group. As before, we may argue that the Poincar\'e group is only a local approximate symmetry and use it to make sense of both the covariance of the quantum fields and the description of particles as local excitations of the (local) vacuum state that we can construct. These difficulties and ambiguities led many researchers to conclude that, ultimately, the notion of elementary particles in curved space-times is meaningless (see, for instance, \cite[Footnote at page $2$]{witten2022does}).

After the previous discussion on the role of symmetries in physical theories and the need to replace groups with groupoids to properly take account of them, it should be obvious that a similar path should be taken when establishing the foundations of relativistic quantum field theories on curved space-times. That is, replacing the usual requirements of Poincar\'e covariance of the quantum fields with covariance with respect to the appropriate kinematical groupoid. As indicated before, in this paper we are going to set the stage of ``groupoids replacing groups'', so that other researchers may join to take up some of the emerging problems.    Several additional conceptual issues will be addressed in the near future as we are presently working on them like the problem of defining covariance for quantum fields on general space-times, the problem of constructing relativistic equations for higher spin theories, etc. (see Sect. \ref{sec:conclusions} for a summary of some of them).

Before ending this introduction, we should mention that in the past a different road was taken to address some of these questions. Geometries of space-times different from Minkowski were obtained by relaxing the structure of the Poincar\'e group. \textit{Bacry} and \textit{Levy-Leblond} characterised the family of relativistic kinematical groups \cite{Bacry-LevyLeblond-Kinematics-1968}, and the corresponding family of space-times defined as homogeneous spaces of them. However, this approach does not allow us to address the problem of describing elementary particles on generic space-times because, again, the restrictive starting point is that symmetries of a space-time are described by means of groups of which they become homogeneous spaces.

We should also mention ideas by \textit{Heller, Pysiak} and \textit{Sasin} \cite{Heller-Pys-Sasin-NoncommutativeUnification-2005, Heller-Pys-Sasin-ConceptualUnification-2007} proposing a model unifying gravity with quantum mechanics that contains insights using groupoids similar to some of the ideas we are going to develop in this paper. However, in our work, we do not pretend to provide any unified model of gravity and quantum mechanics to reach a mathematically and physically meaningful notion of elementary particles; rather, we propose simply an extension of Wigner's program based on the notion of symmetry determined by groupoids and their projective representations.

There are many other works where groupoids are being used in understanding foundational aspects in Physics, for instance in \cite{Blohmann2013} it is shown how the constraint algebra of general relativity is neatly described as the Lie algebroid of a certain groupoid.   However, to the best of our knowledge this work is the first attempt to address the problem of classification of elementary particles beyond the use of Poincar\'e's group.

The paper is organized as follows. As already indicated, Sect. \ref{sec:wigner_classification} will be devoted to succinctly reviewing Wigner's classification of elementary particles on Minkowski space-time, starting with the mathematical theorem by \textit{G.W. Mackey} whose generalization we are seeking. After this, Sect. \ref{Sec: Relativistic groupoids: The Wigner groupoid} will be devoted to the analysis of the notion of symmetry, the introduction of groupoids, and the definition and analysis of the kinematical groupoids that will be relevant for the present work: Poincar\'e and Wigner groupoids. In Section \ref{Sec: Mackey's imprimitivity theorem and the theory of unitary representation of groupoids}, the main ideas leading Wigner to identify elementary particles with irreducible projective representations of the Poincar\'e group will be extended to groupoids in a natural way. Then, the main mathematical theorem, the counterpart of Mackey's theorem that we need to study the theory of projective representations of Lie groupoids, will be established. Finally, in Sect. \ref{Sec: Relativistic particles and relativistic groupoids: Wigner's elementary particles revisited}, the previous ideas will be applied to a generic space-time by computing explicitly the irreducible projective representations of its Wigner groupoid and showing that they produce a classification similar to Wigner's, except for a new class of representations characterized by a magnetic moment $\mu$. The analysis of the physical implications of the new class of representations will be discussed elsewhere. The paper will conclude by summarizing the conclusions and problems that open up once this new approach to symmetry is considered.


\section{Classification of elementary particles on Minkowski space-time revisited}\label{sec:wigner_classification}

In Minkowski space-time $\mathbb{M}^{1,3} \,=\, (\mathbb{R}^4,\,  \eta_0 )$, with $ \eta_0 $ denoting the Minkowski metric tensor with $(1,\,3)$-signature, elementary particles are defined as irreducible projective representations of the connected component of its global isometry group, the restricted Poincar\'e group $\mathcal{P} \,=\, \mathbb{T}^4 \ltimes SO_0(1,3)$, where $SO_0(1,3)$ denotes the restricted Lorentz group or, in other words, the connected component of the identity of $SO(1,3)$.

This identification originates from \textit{Wigner's programme} \cite{Wigner-UnitaryRepresentations-1939}, which recognizes that any physically meaningful notion of a particle must transform consistently under space-time symmetries and, thus, elementary particles will correspond to irreducible projective representations of the kinematical group of the theory (see a detailed discussion in \cref{sec:projective}).

  The modern construction of irreducible projective representations of the Poincar\'e group is achieved by observing that they are in one-to-one correspondence with irreducible unitary representations of its universal double covering $\widetilde{\mathcal{P}} = \mathbb{T}^4 \ltimes SL(2,\mathbb{C})$ \cite{Bargmann-UnitaryRay-1954,simms1968lie}.
Then, Mackey's machine, \cref{Thm: Mackey semidirect}, is applied. This provides the characterization and construction of irreducible unitary representations of regular semi-direct products of groups (see, for instance, \cite{mackey1952}, or \cite[Section 6.6]{Folland-HarmonicAnalysis-2016} for a modern treatment of the subject and references therein) as stated in the following theorem\footnote{In \cref{sec:massless} an extended version of this theorem will be stated, suitable to analyze the classification of massless particles, \cref{thm:non_abelian_extension}.}.

\begin{theorem} (\textsc{Mackey's machine for regular semi-direct products} \cite[Theorem 6.43]{Folland-HarmonicAnalysis-2016}) \label{Thm: Mackey semidirect}
Let \( G = N \ltimes H \) be a second countable locally compact group which is the semi-direct product of an Abelian closed subgroup $N$ and a closed subgroup $H$.
Let $G$ act regularly on $N$, i.e., there is a Borel set in $\widehat{N}$, the Pontryagin dual of $N$, consisting of all characters $ \chi\;:\;\; N \to U(1)$ of $N$, such that it intersects each orbit of $G$ at exactly one point.

  Then, unitary irreducible representations of \(G\) are classified by the following data:
\begin{enumerate}
    \item An orbit \(\mathcal{O}_{\chi_0} \subset \widehat{N}\) under the action of \(H\), where $\chi_0$ is a representative point of the orbit and \(H\) acts on \(\widehat{N}\) by $(h \cdot \chi)(n) := \chi(h^{-1} \cdot n)$ \, for $h \in H$, $n \in N$, and $\chi \in \widehat{N}$;
    \item A unitary irreducible representation \(\Phi_H\;:\;\; H_{\chi_0} \to \mathcal{U}(V_{H_{\chi_0}})\) of the stabilizer subgroup of $\chi_0$:
    \be
    H_{\chi_0} := \{ h \in H \;:\;\; h \cdot \chi_0 = \chi_0 \}\,.
    \ee
\end{enumerate}
Indeed, a unitary irreducible representation of \(G\) is given by the induced representation
    \be
    \Phi_G := \mathrm{Ind}^{G}_{G_{\chi_0}} (\chi_0 \Phi_H)\,,
    \ee
and every irreducible unitary representation of \(G\) is unitarily equivalent to it. Moreover, two irreducible representations $\mathrm{Ind}^{G}_{G_{\chi_0}} (\chi_0 \Phi_H)$ and $\mathrm{Ind}^{G}_{G_{\chi_0}} (\chi_0' \Phi_H')$ are equivalent iff $\chi_0$ and $\chi_0'$ belong to the same orbit, say $\chi_0' = g\cdot \chi_0$, and the representations $h \mapsto \Phi_H(h)$ and $h \mapsto \Phi_H'(ghg^{-1})$ are equivalent.
\end{theorem}

  The induced representation $\mathrm{Ind}^{G}_{G_{\chi_0}} (\chi_0 \Phi_H)$ takes the explicit form:
\[
\bigl( \Phi_G(n, h)\,f \bigr)(h') = \chi_0 \,\big(h'^{-1} \cdot n\big) \cdot f(h^{-1} h') \, , \qquad \forall n \in N, h \in H \, ,
\]
where $f$ belongs to the Hilbert space $\mathcal{H}$
\[
\begin{split}
\mathcal{H} := \Bigl\{\, f\;:\;\; H \to V_H \;:\;\; &f(h_0 h) = \Phi_H(h_0)^{-1} f(h), \;\forall h_0 \in H_{\chi_0},\; h \in H, \text{ and } \\
&\int_{H/H_{\chi_0}} \|f(h)\|^2 \, d\mu(h) < \infty \,\Bigr\}
\end{split}
\]
where \(\mu\) is a quasi-invariant measure on the homogeneous space \(H/H_{\chi_0}\).

\bigskip

  The application of Mackey's machine to the Poincar\'e group provides the standard classification of elementary particles on Minkowski space.

\begin{example}[\textsc{irreducible projective representations of the Poincar\'e group}]
Let \(\widetilde{\mathcal{P}} = \mathbb{T}^4 \ltimes SL(2,\mathbb{C})\) be the universal covering of the Poincar\'e group, where \(\mathbb{T}^4 \cong \mathbb{R}^4\) denotes the group of space-time translations.
In this case \(N = \mathbb{T}^4\), so that \(\widehat{N} \cong \mathbb{R}^4\) via Fourier duality. Each character \(\chi_p\) corresponds to a 4-momentum \(p_\mu\):
    \be
    \chi_p \;:\;\; \mathbb{R}^4 \to U(1) \;:\;\; x \mapsto \chi_p(x) \,=\, e^{i p_\mu x^\mu} \, .
    \ee
    The Lorentz group and its double covering $SL(2, \mathbb{C})$ act on \(\widehat{\mathbb{T}^4} \cong \mathbb{R}^4\) via the standard representation. Such action preserves the invariant $p^2 \,=\, p_\mu p^\mu$.
    Thus, orbits of the standard action of $SL(2,\mathbb{C})$ upon $\widehat{\mathbb{T}^4}$ are labelled by the values of $p^2$.
    In particular, they can be collected in the following three families:
    \begin{enumerate}
        \item Orbits for which \(p^\mu p_\mu \,=:\, m^2 > 0\). We will consider the future component of the mass shell $H_m^+ = \{ p^2 =m^2> 0, p_0 > 0 \} $.
        A representative of each of these orbits is $\chi_0 \,=\, (m,\,0,\,0,\,0)$, whose little group is $SU(2)$, for which unitary irreducible representations are labelled by a semi-integer $s \,=\, \frac{1}{2} n$, for $n \in \mathbb{N}_0$, physically representing the \textit{spin} of the particle of \textit{mass} $m$.
        \item Orbits for which $p^\mu p_\mu = 0$. We will consider the future component of the light cone $C^+ = \{ p^2 = 0, p_0 > 0 \}$.
        A representative of this orbit is $\chi_0 \,=\, (E,\,0,\,0, E)$ for $E > 0$, whose little group is a double cover of the Euclidean group $E(2)$, for which unitary irreducible representations are one-dimensional and labelled by a half-integer number $\lambda \in \mathbb{Z}/ 2$, physically representing the \textit{helicity} of the \textit{massless} particle; or infinite-dimensional and labelled by a continuous parameter $\rho$. Such representations are called continuous spin representations and are rejected for physical reasons, even if there is a renewed interest in their study \cite{Gracia-Bondia-2018}.
        \item Orbits for which $p^\mu p_\mu < 0$. These orbits are usually neglected or associated with tachyons.
    \end{enumerate}
    Orbits corresponding to the past component of the mass shell or the light cone are understood as antiparticles of the corresponding ones.
Therefore, induced representations of $SL(2,\mathbb{C})$ on $\widetilde{\mathcal{P}}$, namely, irreducible unitary representations of \(\widetilde{\mathcal{P}}\), are: Massive particles with mass $m > 0$ and spin $s$; massless particles with helicity $\lambda$.
\end{example}


\section{Relativistic groupoids: The Wigner groupoid}
\label{Sec: Relativistic groupoids: The Wigner groupoid}


\subsection{Symmetries and groupoids}\label{sec:groupoids_symmetry}

Generic space-times do not have nontrivial groups of isometries.
Hence, a generic space-time $\mathscr{M}$ is not the homogeneous space of a group of isometries acting on it.
Thus, there is no a priori reason to single out the Poincar\'e group as a fundamental symmetry.
Indeed, the global isometry group \(G{(\mathscr{M},\,\eta)}\) is, by definition, the set of all diffeomorphisms $\phi\colon \mathscr{M} \,\to\, \mathscr{M}$,
satisfying
\be
\phi^*\eta \,=\, \eta \,.
\ee

  For a generic choice of smooth metric \(\eta\), the only diffeomorphism that preserves \(\eta\) is the identity map, so \(G{(\mathscr{M},\,\eta)}\) is trivial.
  Consequently, any definition of an elementary particle built on the notion of the symmetry provided by the global isometry group of the given space-time breaks down, motivating the shift to a notion of symmetry structure that remains meaningful in arbitrary geometries.

  The main idea of the present paper is that the geometric structure capable of encoding such a generalized notion of symmetry is that of a \textit{groupoid}.

\medskip

\noindent\textbf{A primer on Groupoids.} For the convenience of the reader, we will establish a few definitions and notations that will be used throughout the paper.

  A groupoid consists of a set of objects $x, y,... \in \Omega$, a set of morphisms $\alpha, \beta, ... \in \mathscr{G}$, and two maps $s,t \colon \mathscr{G} \to \Omega$, called the source and target maps respectively, such that if $s(\alpha ) = x$ and $t(\alpha) = y$, then we will denote the morphism $\alpha$ as $\alpha \colon x \to y$.
Additionally, there is a partial composition law $\circ \colon \mathscr{G}^{(2)} \to \mathscr{G}$, where $\mathscr{G}^{(2)} = \{ (\alpha, \beta) \in \mathscr{G} \times \mathscr{G} \mid s(\alpha) = t(\beta) \}$ is the set of composable morphisms. The composition law $\circ$ is associative, and, for any object $x\in \Omega$, there is a morphism $1_x \colon x \to x$, such that $\alpha \circ 1_{s(\alpha)} = \alpha = 1_{t(\alpha)} \circ \alpha$.
Finally, for any $\alpha \colon x \to y$, there exists $\alpha^{-1} \colon y \to x$, such that $\alpha^{-1} \circ \alpha = 1_x$, $\alpha \circ \alpha^{-1} = 1_y$.
The morphisms $1_x$ are called units, and the groupoid will often be denoted as $\mathscr{G} \rightrightarrows \Omega$ (see, for instance, \cite{Ibort-Rodriguez-IntroGroupoids-2019} for the basic definitions of groupoids and their role as generalized symmetries).

  Given an object $x\in \Omega$ of the groupoid $\mathscr{G} \rightrightarrows \Omega$, the orbit through $x$ will be defined as:
\be
O_x \,=\, \{ y \,\in\, \Omega \;:\;\; \exists \,\alpha\,:\, x \to y \} \,.
\ee
The set of orbits of a groupoid will be denoted as $\Omega/\mathscr{G}$ and defines a partition of $\Omega$, $\Omega = \sqcup_{O \in \Omega/\mathscr{G}} O$. A groupoid will be called transitive if it has just one orbit or, equivalently, given two objects $x,y\in \Omega$, there is always a morphism $\alpha \colon x \to y$.

  Given $x \in \Omega$, the collection of all morphisms $\alpha \colon x \to y$, $y\in \Omega$, that is $s^{-1}(x)$, will be denoted as $\mathscr{G}_x$, and similarly, $\mathscr{G}^x$ will denote the set $t^{-1}(x)$. The set $\mathscr{G}_x \cap \mathscr{G}^x$, that is, the set of all morphisms such that $\alpha \colon x \to x$, forms a group called the isotropy group of the groupoid $\mathscr{G}$ at $x$ and denoted by $\mathscr{G}(x)$ or $G_x$.
It is easy to check that if $x,y$ belong to the same orbit, then their corresponding isotropy groups are isomorphic.

  Given any subset $S \subset \Omega$ of the space of objects of $\mathscr{G}$, we define the restriction of the groupoid $\mathscr{G}$ to $S$, say $\mathscr{G}_S$, as $\mathscr{G}_S = \{ \alpha\,:\, x \to y \in \mathscr{G}\,, \;\;\text{s.t.}\; x,y \in S \}$.
Note that $\mathscr{G}_S$ is a groupoid with space of objects $S$ or, using the previous notation, $\mathscr{G}_S \rightrightarrows S$.

Notably, the disjoint union of groupoids is again a groupoid; that is, given groupoids $\mathscr{G} \rightrightarrows \Omega$, $\mathscr{G}' \rightrightarrows \Omega'$, then $\mathscr{G} \sqcup \mathscr{G}' \rightrightarrows \Omega\sqcup \Omega'$ is again a groupoid with the obvious source and target maps and partial composition law.
Then, it is a simple exercise to check that a groupoid is the disjoint union of the restrictions to its orbits, that is:
\begin{equation}\label{eq:structure}
\mathscr{G} = \bigsqcup_{O \in \Omega/\mathscr{G}} \mathscr{G}_O \, .
\end{equation}
Each one of the groupoids $\mathscr{G}_O$ will be called a component of the groupoid $\mathscr{G}$
(see, for instance, \cite{Ibort-Rodriguez-IntroGroupoids-2019} for the basic structure theorems of groupoids).

  A Lie groupoid is a groupoid $\mathscr{G} \rightrightarrows \Omega$ such that both $\mathscr{G}$ and $\Omega$ are smooth manifolds, the source and target maps are smooth submersions, the natural inclusion map $i \colon \Omega \to \mathscr{G}$, given by $i(x) = 1_x$, is an immersion, and both the partial composition law $\circ$ and the inversion map $\tau \colon \mathscr{G} \to \mathscr{G}$, $\tau (\alpha) = \alpha^{-1}$, are smooth maps (see, for instance, \cite{Mackenzie-LieGroupoids-2005} for a detailed description of the theory of Lie groupoids).

  If $\mathscr{G} \rightrightarrows \Omega$ is a Lie groupoid, then the restrictions of $\mathscr{G}$ to the connected components (as a topological space) of the manifold $\Omega$ are again Lie groupoids, so that $\mathscr{G} = \sqcup_{a\in \pi_0(\Omega)} \mathscr{G}_{\Omega_a}$, where $\{ \Omega_a \}$ denote the connected components of $\Omega$. In what follows we will assume that $\Omega$ is a connected manifold.

  The orbits $O$ of the Lie groupoid $\mathscr{G} \rightrightarrows \Omega$ are smooth submanifolds and they define a generalized foliation of $\Omega$.
A typical example of this situation happens when $G$ is a Lie group acting smoothly on the manifold $\Omega$ and $\mathscr{G} = G \times \Omega \rightrightarrows \Omega$ is the action groupoid associated to such action with source and target maps $s (g,x) = x$, $t(g,x) = gx$, $g \in G$, $x\in \Omega$.
The action groupoid $\mathscr{G} = G \times M \rightrightarrows M$ is a Lie groupoid.
Moreover, the orbit $O_x$ of the action groupoid passing through the element $x\in \Omega$ coincides with the orbit $Gx$ of the action of the group $G$ on $\Omega$ passing through $x$, and the space of orbits $\Omega/\mathscr{G}$ is the quotient space $\Omega/G$ of $\Omega$ under the action of the group $G$.

The restriction of a Lie groupoid $\mathscr{G}$ to any of its orbits $O$ is again a Lie groupoid $\mathscr{G}_O$, and the right hand side of (\ref{eq:structure}) becomes a smooth manifold whose connected components are the groupoids $\mathscr{G}_O$, while the left hand side is a smooth manifold which is connected as a topological space.
Hence, the decomposition of a groupoid as the disjoint union of its components does not reconstruct the smooth structure of the overall groupoid, and (\ref{eq:structure}) is not an isomorphism of Lie groupoids.

  We end this summary of basic notions on groupoids by observing that if $G$ is a Lie group acting on a manifold $\Omega$, then the action groupoid $\mathscr{G} = G \times \Omega \rightrightarrows \Omega$ captures the structure of the action of the group $G$ on $\Omega$; hence we can think about the properties of geometrical objects on $\Omega$ invariant under the action of $G$ in terms of the properties of the action groupoid $\mathscr{G}$.
  
For instance, if we consider the standard action of the Poincar\'e group $\mathcal{P}$ on Minkowski space $\mathbb{M}^{1,3}$, then the action groupoid $\mathscr{G}(\mathbb{M}^{1,3}) = \mathcal{P} \times \mathbb{M}^{1,3}$ is a transitive Lie groupoid of dimension 14 capturing all the properties of such action. 

Moreover, if we pick a submanifold $S \subset \Omega$, the restriction groupoid $\mathscr{G}_S \rightrightarrows S$ is still a Lie groupoid with the space of objects being the submanifold $S$, while, in general, there is no Lie group $G'$ such that $\mathscr{G}_S \rightrightarrows S$ is the action groupoid corresponding to the action of $G'$ on $S$. Consider, for instance, any open set $U \subset \mathbb{M}^{1,3}$. Then, the restriction of the action groupoid $\mathscr{G}(\mathbb{M}^{1,3})$ to $U$ will consist of all pairs $((\Lambda,a), x)$, $x \in \mathbb{M}^{1,3}$, $(\Lambda,a)\in \mathcal{P}$, such that $x, \Lambda x +a \in U$ (which is not empty), but for a generic $U$, there is no group of isometries leaving $U$ invariant. We will say that the restriction groupoid $\mathscr{G}(\mathbb{M}^{1,3})_U$ describes the ``intrinsic'' symmetry of the domain $U$ despite the fact that there is no symmetry group of isometries acting on $U$\footnote{We may argue that the group of diffeomorphisms of $M$ preserving $U$ is the ``ultimate'' group of symmetries of $U$. Indeed, this group is related to the group of bisections of the restriction of the action groupoid to $U$. We will not use this approach in this discussion, leaving it for further analysis.}. This simple observation contains the essence of the main idea that we are developing in this paper: ``even if there are no groups describing the symmetry properties of a given system there could still exist a groupoid doing it''.

\vspace{1em}

We end this section by turning to its title and moving beyond Weyl's principle, which states that symmetries are implemented via groups of transformations, to what could be called the \textit{Weyl-Weinstein principle} \cite{Weinstein-GroupoidsSymmetry-1996}, asserting that symmetries are implemented via groupoids.
An extension of such a principle, inspired by the groupoidal description of quantum mechanical systems where groupoids appearing as symmetries of quantum systems are called ``symmetroids'', was introduced in \cite{Ciaglia+-DynamicalMaps-2021, Ciaglia-DiCosmo-Ibort-Marmo-Schiavone-Symmetroids-2021}, even though this refined notion of symmetry will not be needed in the present context.


\subsection{The Poincar\'e groupoid of a space-time}

In this section we will show that we can associate to any space-time $(\mathscr{M},\, \eta)$ various groupoids that capture its ``intrinsic'' symmetry regardless of whether there are global isometries or not. We begin by describing a Lie groupoid associated with $(\mathscr{M},\, \eta)$:
\begin{eqnarray}
\mathsf{Poincare}{(\mathscr{M},\, \eta)} &=& \bigl\{\, \left(y,\,T_{yx},\,x\right) \,\mid\, x,\, y \in \mathscr{M}\,, \nonumber \\ && T_{yx}\in \mathrm{Lin}\left(\mathbf{T}_x \mathscr{M},\,\mathbf{T}_y \mathscr{M}\right) \,,\;\; T_{yx}^* \eta_y \,=\, \eta_x \,\bigr\} \,,
\end{eqnarray}
to which we will refer as the \textit{Poincar\'e groupoid} of the space-time $\mathscr{M}$\footnote{This groupoid is also referred to as the groupoid of frames of $\mathscr{M}$ \cite{Mackenzie-LieGroupoids-2005}.}.
Its space of objects is $\mathscr{M}$, and the space of morphisms is the set of linear maps  $T_{yx}$ between pairs of tangent spaces to $\mathscr{M}$ preserving the metric structure $\eta$.
The source and the target maps are given by:
\begin{align}
s \;:\;\; \mathsf{Poincare}{(\mathscr{M},\,\eta)} \to \mathscr{M} \;:\;\; \left(y,\,T_{yx},\,x\right) \mapsto s \left(y,\,T_{yx},\,x\right) \,=\, x \,, \\
t \;:\;\; \mathsf{Poincare}{(\mathscr{M},\,\eta)} \to \mathscr{M} \;:\;\; \left(y,\,T_{yx},\,x\right) \mapsto t \left(y,\,T_{yx},\,x\right) \,=\, y \, ,
\end{align}
and the partial composition law of $\mathsf{Poincare}{(\mathscr{M},\, \eta)}$ reads:
\be
(z,\,T_{zy},\, y) \circ (y,\,T_{yx},\,x) \,=\, (z,\,T_{zy}\circ T_{yx},\,x) \,.
\ee
Since $\mathbf{T}_x \mathscr{M}$ and $\mathbf{T}_y \mathscr{M}$ have the same dimension, and since both $\eta_y$ and $\eta_x$ are non-degenerate with the same signature, there always exists at least one linear isometry between $\mathbf{T}_x \mathscr{M}$ and $\mathbf{T}_y \mathscr{M}$ for any pair of $x,\, y \in \mathscr{M}$, for instance, the linear map mapping a given orthonormal basis of $\mathbf{T}_x \mathscr{M}$ w.r.t. $\eta_x$ to another orthonormal basis of $\mathbf{T}_y \mathscr{M}$ w.r.t. $\eta_y$.
Thus, $\mathsf{Poincare}{(\mathscr{M},\, \eta)}$ is transitive.

  Moreover, $\mathsf{Poincare}{(\mathscr{M},\, \eta)}$ is a Lie groupoid whose smooth structure is constructed by using the smooth structure of $\mathscr{M}$, which provides local trivializations of $\mathbf{T}\mathscr{M}$ determining local expressions for the linear transformations $T_{yx}$.
Specifically, consider an atlas $(U_j,\,\varphi_j)_{j \in \mathscr{J}}$ on $\mathscr{M}$ ($\mathscr{J}$ denoting an index set)
\be
\varphi_j \;:\;\; \mathscr{M} \supset U_j \to \mathbb{R}^n \;:\;\; m \mapsto \varphi_j(m) \,=\, x^\mu \,, \;\; \mu=1,...,n \,.
\ee
Given any open set $U_j$ of the atlas above, consider the local trivialization of $\mathbf{T}\mathscr{M}$ defined by it:
\be \label{Eq: trivialization tangent bundle}
\psi_j \;:\;\; \mathbf{T}U_j \to U_j \times \mathbb{R}^n \;:\;\; v_m \mapsto (x^\mu,\, v^\mu)\,, \;\; \mu,\,I=1,...,n,\,,
\ee
and any local orthonormal basis: $\left\{\, e_I(x) \,\right\}_{I=1,...,n}$, i.e., a local basis of tangent vectors on $U_j$ satisfying:
\be \label{Eq: definition tetrads}
\eta_{\mu \nu}(x) e^\mu_I(x) e^\nu_J(x) \,=\, \eta_{IJ} \,,
\ee
for $\eta_{IJ}$ the Minkowski metric on $\mathbb{R}^n$.

  Local charts on $\mathsf{Poincare}{(\mathscr{M},\,\eta)}$ are given by the open sets:
\be
\tilde{U}_{kj} \,=\, U_k \times U_{\mathrm{GL}_{kj}} \times U_j \,, \;\; j,\,k \in \mathscr{J} \,,
\ee
where $U_j$ and $U_k$ are open sets of the local chart on $\mathscr{M}$ constructed above, while
\be
U_{\mathrm{GL}_{kj}} \,=\, \left\{\, T_{yx} \;:\;\; \mathbf{T}_x U_k \to \mathbf{T}_y U_j \;:\;\; x \in U_k,\, y \in U_j,\;\; {T_{yx}}^*\eta_y \,=\, \eta_x \,\right\} \,,
\ee
is the subset of linear operators from tangent spaces to points in $U_k$ to tangent spaces to points in $U_j$ preserving the metric.
Coordinates on such open sets are given by:
\be
\tilde{\psi} \;:\;\; \tilde{U}_{kj} \to \mathbb{R}^n \times \mathbb{R}^{n^2} \times \mathbb{R}^n \;:\;\; p \mapsto \tilde{\psi}(p) \,=\, (y^\mu,\, \Lambda^I_J(y,\,x),\, x^\mu) \,,
\ee
where
\be
T_{yx} e_I(x) \,=\, \Lambda^J_I(y,\,x) e_J(y) \,.
\ee
Thus, we conclude that $\mathsf{Poincare}{(\mathscr{M},\,\eta)}$ is a smooth manifold of dimension $2n + \frac{n(n-1)}{2}$, thus if $n = 4$ we get that the Poincar\'e groupoid has dimension 14 and, topologically, it is connected provided that $\mathscr{M}$ is connected.

  Note that if $(\mathscr{M},\, \eta)$ admits a global isometry group $G{(\mathscr{M},\,\eta)}$, the linear maps
\be \label{Eq: iso symmetry groupoid isometry group}
T_{yx} \,=\, T\Phi\mid_{T_x\mathscr{M}} \,,
\ee
where $\Phi\colon \mathscr{M} \to \mathscr{M}$ is an element of $G{(\mathscr{M},\,\eta)}$ such that $\Phi(x) = y$, belong to $\mathsf{Poincare}{(\mathscr{M},\, \eta)}$.
On the other hand, when $G{(\mathscr{M},\, \eta)}$ is empty, $\mathsf{Poincare}{(\mathscr{M},\, \eta)}$ is still non-trivial.
Indeed, as argued above, non-trivial linear maps \(T_{yx}\) between every pair of tangent spaces always exist, even if they may not integrate to globally defined diffeomorphisms.
Therefore, $\mathsf{Poincare}{(\mathscr{M},\, \eta)}$ encompasses all the information contained in the global isometry group (when it exists) and generally represents a richer geometric structure by also capturing the ``local'' symmetries of $\eta$.

  The isotropy group of $\mathsf{Poincare}{(\mathscr{M},\,\eta)}$ at $x \in \mathscr{M}$ is the set of linear transformations of $\mathbf{T}_x \mathscr{M}$ preserving $\eta_x$:
\be
G_x \,=\, \left\{\, T_{xx} \in \mathrm{Hom}(\mathbf{T}_x \mathscr{M}) \;:\;\; {T_{xx}}^*\eta_x \,=\, \eta_x \,\right\} \,.
\ee
Since $\eta$ has been assumed to have $(1,\,3)$-signature, $G_x$ is by definition the orthogonal group $O(1,\,3)$ for any $x \in \mathscr{M}$.
The fact that all isotropy groups are isomorphic follows from the transitivity of \(\mathsf{Poincare}{(\mathscr{M}, \eta)}\).
Specifically, for any two points \(x\) and \(y\), the isotropy groups \(G_x\) and \(G_y\) are related by
\[
G_x = {T_{yx}}^{-1}\, G_y\, T_{yx}\,, \;\;\; \text{for some }T_{yx} \in \mathrm{Hom}(\mathbf{T}_x\mathscr{M},\,\mathbf{T}_y \mathscr{M}) \,,
\]
where \(T_{yx}\) always exists due to the transitivity.

\begin{example}[\textsc{Minkowski metric}]
Consider $(\mathscr{M},\,\eta)$ to be Minkowski space-time, say $(\mathbb{M}^{1,3},\, \eta_0 )$.
In this case:
\begin{eqnarray}
\mathsf{Poincare}{(\mathbb{M}^{1,3},\,  \eta_0 )} &=& \bigl\{\, \left(y,\,T_{yx},\,x\right) \,\mid\,\, x,\, y \in \mathbb{M}^{1,3}\,, \nonumber \\ && T_{yx}\in \mathrm{Lin}\left(\mathbf{T}_x \mathbb{M}^{1,3},\,\mathbf{T}_y \mathbb{M}^{1,3}\right) \,,\;\; T_{yx}^* { \eta_0 }_y = { \eta_0 }_x \,\bigr\} \,.
\end{eqnarray}
The object space is an affine space $\mathbb{M}^{1,3}$ admitting a global chart with the associated system of coordinates $\left\{\, x^\mu \,\right\}_{\mu=0}^3$.
Thus $\mathbf{T}\mathbb{M}^{1,3}$ admits a global trivialization as
\be
\mathbf{T}\mathbb{M}^{1,3} \,=\, \mathbb{M}^{1,3} \times \mathbb{R}^4 \,,
\ee
providing the global decomposition of $\mathsf{Poincare}{(\mathbb{M}^{1,3},\, \eta_0 )}$
\be
\mathsf{Poincare}{(\mathbb{M}^{1,3},\, \eta_0 )} \,=\, \mathbb{M}^{1,3} \times \mathrm{O}(1,\,3) \times \mathbb{M}^{1,3} \,,
\ee
equipped with the global system of coordinates:
\be
\left\{\, x^\mu,\, \Lambda^\mu_\nu(y,\,x),\, y^\mu \,\right\}_{\mu,\,\nu=0}^3 \,,
\ee
where $\Lambda^\mu_\nu(y,\,x)$ is the matrix representing the element of $\mathrm{O}(1,\,3)$ connecting $x$ and $y$, written in the global orthonormal basis $\{ \partial /\partial x^\mu \}$.
  It is relevant to point out that the Poincar\'e groupoid of Minkowski space is isomorphic to the action groupoid of the action of the Poincar\'e group on Minkowski space, that is, with the notations above:
\begin{equation}\label{eq:Poincare_decomp}
\mathsf{Poincare}{(\mathbb{M}^{1,3},\,  \eta_0 )} = \mathscr{G}(\mathbb{M}^{1,3}) = \mathcal{P} \times \mathbb{M}^{1,3} \, .
\end{equation}
Indeed, the map $(y,T_{yx}, x) \mapsto ((\Lambda, a), x)$, with $y = x + a$, $T_{yx}(\partial/\partial x^\mu) = \Lambda_{\mu}^\nu \partial/\partial x^\nu$, defines the desired identification.
This reinforces the discussion at the end of \cref{sec:groupoids_symmetry}, showing that groupoids capture the notion of symmetry defined by group actions.
\end{example}

\begin{example}[\textsc{Schwarzschild metric}]
Consider the Schwarzschild space-time \((\mathbb{M}^{1,3}_{0},\,\eta_S)\), where $\mathbb{M}^{1,3}_0$ is the vector space $\mathbb{M}^{1,3} \backslash \left\{\, 0 \,\right\}$, and $\eta_S$ is the Schwarzschild metric that in standard Schwarzschild coordinates reads
\[
\eta_S \,=\, \left(1-\frac{2M}{r}\right) \dd t \otimes \dd t - \left(1-\frac{2M}{r}\right)^{-1} \dd r \otimes \dd r - r^2\left(\, \dd \theta \otimes \dd \theta + \sin^2\theta\, \dd \phi \otimes \dd \phi \,\right)\,,
\]
with \(M\) the mass parameter and the hypersurface defined by $r \,=\, 2M$ being the event horizon.
In these coordinates, the metric is singular at \(r=2M\).
For this reason, one usually introduces Kruskal coordinates, which provide a maximal analytic extension of the Schwarzschild solution and remove the coordinate singularity at the horizon \cite{Wald-GeneralRelativity-1984}.
Kruskal coordinates \((U,\,V,\,\theta,\,\phi)\) are defined by the transformations
\[
\begin{aligned}
U &= -\left(\frac{r}{2M}-1\right)^{1/2} e^{r/4M} \sinh\left(\frac{t}{4M}\right),\\[1mm]
V &= \left(\frac{r}{2M}-1\right)^{1/2} e^{r/4M} \cosh\left(\frac{t}{4M}\right),
\end{aligned}
\]
for \(r>2M\) (with the appropriate analytic continuation for \(r<2M\)).
In these coordinates the metric becomes
\[
\eta_S \,=\, -\frac{32M^3}{r}\,e^{-r/2M}\,\dd U \otimes_S \dd V - r^2\left(\,\dd \theta \otimes \dd \theta + \sin^2\theta\, \dd \phi \otimes \dd \phi \,\right)\,,
\]
where $\otimes_S$ is the symmetrized tensor product, and where \(r\) is implicitly defined by
\[
UV \,=\, \left(1-\frac{r}{2M}\right)e^{r/2M}\,.
\]
This form is manifestly regular at the event horizon (\(r=2M\), corresponding to \(U=0\) or \(V=0\)) and covers both the exterior (\(r>2M\)) and interior (\(0<r<2M\)) regions (note that the genuine curvature singularity at \(r=0\) is not removed).

  In a given chart defined in Kruskal coordinates one can introduce a local trivialization of the tangent bundle \(\mathbf{T}\mathscr{M}\) by defining (up to a local Lorentz transformation at each fixed tangent space) a local tetrad basis
\be
\begin{aligned}
e_0 &\,=\, \sqrt{\frac{r}{32M^3}}\frac{e^{r/4M}}{\sqrt{2}}\left(\frac{\partial}{\partial V} + \frac{\partial}{\partial U}\right),\\[1mm]
e_1 &\,=\, \sqrt{\frac{r}{32M^3}}\frac{e^{r/4M}}{\sqrt{2}}\left(\frac{\partial}{\partial V} - \frac{\partial}{\partial U}\right),\\[1mm]
e_2 &\,=\, \frac{1}{r}\,\frac{\partial}{\partial \theta},\\[1mm]
e_3 &\,=\, \frac{1}{r\sin\theta}\,\frac{\partial}{\partial \phi}.
\end{aligned}
\ee
Its dual, namely the basis of tetrad one-forms, reads
\be \label{Eq: tetrad one-forms Schwarzschild}
\begin{aligned}
e^0 &\,=\, \sqrt{\frac{16M^3}{r}}\,e^{-r/4M}\,\bigl(\dd V + \dd U\bigr),\\[1mm]
e^1 &\,=\, \sqrt{\frac{16M^3}{r}}\,e^{-r/4M}\,\bigl(\dd V - \dd U\bigr),\\[1mm]
e^2 &\,=\, r\,\dd \theta,\\[1mm]
e^3 &\,=\, r\sin\theta\,\dd \phi.
\end{aligned}
\ee
This choice ensures that the metric locally takes the Minkowskian form
\[
\eta \,=\, \eta_{IJ}\,e^I \otimes e^J\,,
\]
with \(\eta_{IJ}\) the Minkowski metric on \(\mathbb{R}^4\).

  Consider an atlas on \(\mathbb{M}^{1,3}_0\) whose charts are defined in Kruskal coordinates.
In each chart, the tangent bundle is locally trivialized by the tetrad fields defined above.
Then, for two charts \(U_k\) and \(U_j\), an open set of the groupoid is given by
\[
\tilde{U}_{kj} \,=\, U_k \times U_{\mathrm{GL}_{kj}} \times U_j\,,
\]
where
\[
U_{\mathrm{GL}_{kj}} \,=\, \Bigl\{\, T_{yx}\;:\; \mathbf{T}_x U_j \to \mathbf{T}_y U_k \;:\; {T_{yx}}^*\eta_y=\eta_x \,\Bigr\}\,.
\]
Local coordinates on \(\tilde{U}_{kj}\) are provided by
\[
\tilde{\psi} \;:\; \tilde{U}_{kj}\to \mathbb{R}^4 \times \mathbb{R}^{16} \times \mathbb{R}^4\;:\; p \mapsto \tilde{\psi}(p) \,=\, (x^\mu,\, \Lambda^I_J,\, y^\mu)\,,
\]
with the relation
\[
T_{yx}\, e_I(x) \,=\, \Lambda^J_I(x,\,y)\, e_J(y)\,,
\]
for $\Lambda^J_I(x,\,y)$ belonging to $O(1,\,3)$.

Thus the Poincar\'e groupoid is well-defined even at the horizon (in Kruskal coordinates), whereas the isometry group is often only considered in the static patch.
\end{example}

  As will be proven in \cref{Sec: Mackey's imprimitivity theorem and the theory of unitary representation of groupoids}, Thm. \ref{Thm: Mackey theorem transitive groupoids},  projective irreducible representations of a transitive Lie groupoid are in one-to-one correspondence with projective representations of its isotropy group.
Therefore, the Poincar\'e groupoid is not the appropriate framework for classifying elementary particles because its irreducible projective representations will be determined by irreducible projective representations of the Lorentz group, which do not provide the quantum numbers of Wigner's classification.


\subsection{The Wigner groupoid of a space-time}\label{sec:wigner}

To construct the appropriate representations needed to classify elementary quantum systems, we introduce a richer groupoid, referred to as the \textit{Wigner groupoid}, and defined as:
\begin{equation}\label{eq:wigner}
\begin{split}
\mathsf{Wigner}{(\mathscr{M},\, \eta)} \,=\,\bigl\{\, (p_y,\, T_{yx},\, p_x) \mid \, \, &p_x \in  \mathbf{T}^*_x \mathscr{M},\, p_y \in  \mathbf{T}^*_y \mathscr{M} ,\,  \\  T_{yx} \in \mathrm{Hom}(\mathbf{T}_x \mathscr{M},\, \mathbf{T}_y \mathscr{M}) \, , \;
&{T_{yx}}^* p_y \,=\, p_x,\, \text{and } {T_{yx}}^*\eta_y \,=\, \eta_x \,\bigr\} \,.
\end{split}
\end{equation}
Morphisms in $\mathsf{Wigner}{(\mathscr{M},, \eta)}$ are linear isometries between pairs of tangent spaces that preserve the metric $\eta$, while the space of objects is now given by the phase space, the cotangent bundle $ \mathbf{T}^* \mathscr{M}$ of $\mathscr{M}$, with coordinates $\left\{\, x^\mu,\,p_\mu \,\right\}_{\mu = 1,...,n}$.
The source and the target maps read:
\begin{align}
s :\; \mathsf{Wigner}{(\mathscr{M},\, \eta)} \to  \mathbf{T}^* \mathscr{M}\, ; \,\, (p_y,\, T_{yx},\,p_x) \mapsto s (p_y,\, T_{yx},\, p_x) = (x,\, p_x) \, , \\
t :\; \mathsf{Wigner}{(\mathscr{M},\, \eta)} \to  \mathbf{T}^* \mathscr{M} \, ; \,\, (p_y,\, T_{yx},\,p_x) \mapsto t  (p_y,\, T_{yx},\, p_x) = (y,\, p_y) \,.
\end{align}
The composition rule reads, again,
\be
(p_z,\, T_{zy},\, p_y) \circ (p_y,\, T_{yx},\, p_x) \,=\, (p_z,\, T_{zy} \circ T_{yx},\, p_x) \,.
\ee
The Wigner groupoid is again a Lie groupoid whose smooth structure is provided analogously to the Poincar\'e groupoid, however $\mathsf{Wigner}{(\mathscr{M},\, \eta)}$ is no more transitive.
Indeed, since morphisms in $\mathsf{Wigner}{(\mathscr{M},\, \eta)}$ preserve $\eta$, they preserve, in particular, the norm of each covector $p_x \in \mathbf{T}_x \mathscr{M}$, that is, given a morphism $(p_y,\,T_{yx},\, p_x)$ in $\mathsf{Wigner}{(\mathscr{M},\,\eta)}$ we get:
\be
p_x^2 \,=\, {\eta^{-1}}_x(p_x,\,p_x) \,=\, {\eta^{-1}}_x ({T_{yx}}^\star p_y,\,{T_{yx}}^\star p_y) \,=\, {\eta^{-1}}_y(p_y,\,p_y) \,=\, p_y^2 \,,
\ee
where $\eta^{-1}$ denote the contravariant metric tensor
\be
\eta^{-1} \,=\, {\eta^{-1}}^{\mu \nu} \frac{\de}{\de x^\mu} \otimes \frac{\de}{\de x^\nu} \,,
\ee
whose coefficients ${\eta^{-1}}^{\mu \nu}$ obey:
\be
{\eta^{-1}}^{\mu \nu} \eta_{\nu \sigma} \,=\, \delta^\mu_\sigma \,.
\ee
This means that, given two points of $ \mathbf{T}^* \mathscr{M}$ with covectors of different norms, they can not be connected by any morphism of $\mathsf{Wigner}{(\mathscr{M},\,\eta)}$.
Thus, $ \mathbf{T}^* \mathscr{M}$ is foliated in an infinite number of orbits parametrized by the values of $p^2$.
In particular, they can be collected into three families similar to those appearing in studying unitary irreducible representation of Poincar\'e group, see Fig. \ref{fig:orbits} below.

\begin{proposition}\label{prop:foliation}
Given a space-time $\mathscr{M}$, its Wigner groupoid $\mathsf{Wigner}(\mathscr{M})$, is not transitive and its orbits are given by the following list:
\begin{enumerate}
    \item (Massive particles\footnote{As it will be shown in \cref{Sec: Relativistic particles and relativistic groupoids: Wigner's elementary particles revisited}, these orbits will give rise to representations of the groupoid that could be properly understood as ``massive'' particles. The same observation applies to the next item in the list.}:) $\mathcal{O}_{m,+} = \{ (x,p) \in \mathbf{T}^*\mathscr{M} \mid  p^2 = m^2\,>\, 0, m \in \mathbb{R}_+ , p_0 > 0 \}$;
    \item (Massless particles:) $\mathcal{O}_{0,+} = \{ (x,p \in \mathbf{T}^*\mathscr{M} \mid p^2 \,=\, 0\, , p_0 \,>\, 0 \}$.
\end{enumerate}
In addition to these orbits there is the zero orbit, that is $\mathcal{O}_{0} = \{ (x, 0) \in \mathbf{T}^*\mathscr{M} \}$, the orbit corresponding to tachyonic particles $\mathcal{O}_{\pm im} = \{ (x,p) \in \mathbf{T}^*\mathscr{M}  \mid p^2 = -m^2\, , m \in \mathbb{R}_+  \}$, and, finally, the components of massive and massless particles in the past component of the mass hyperboloid and light cone respectively. 
\end{proposition}
As it will be discussed in \cref{Sec: Relativistic particles and relativistic groupoids: Wigner's elementary particles revisited}, this foliation will be instrumental in recovering Wigner's classification of elementary particles in curved space-times.

\begin{figure}
\centerline{\includegraphics[width=12cm]{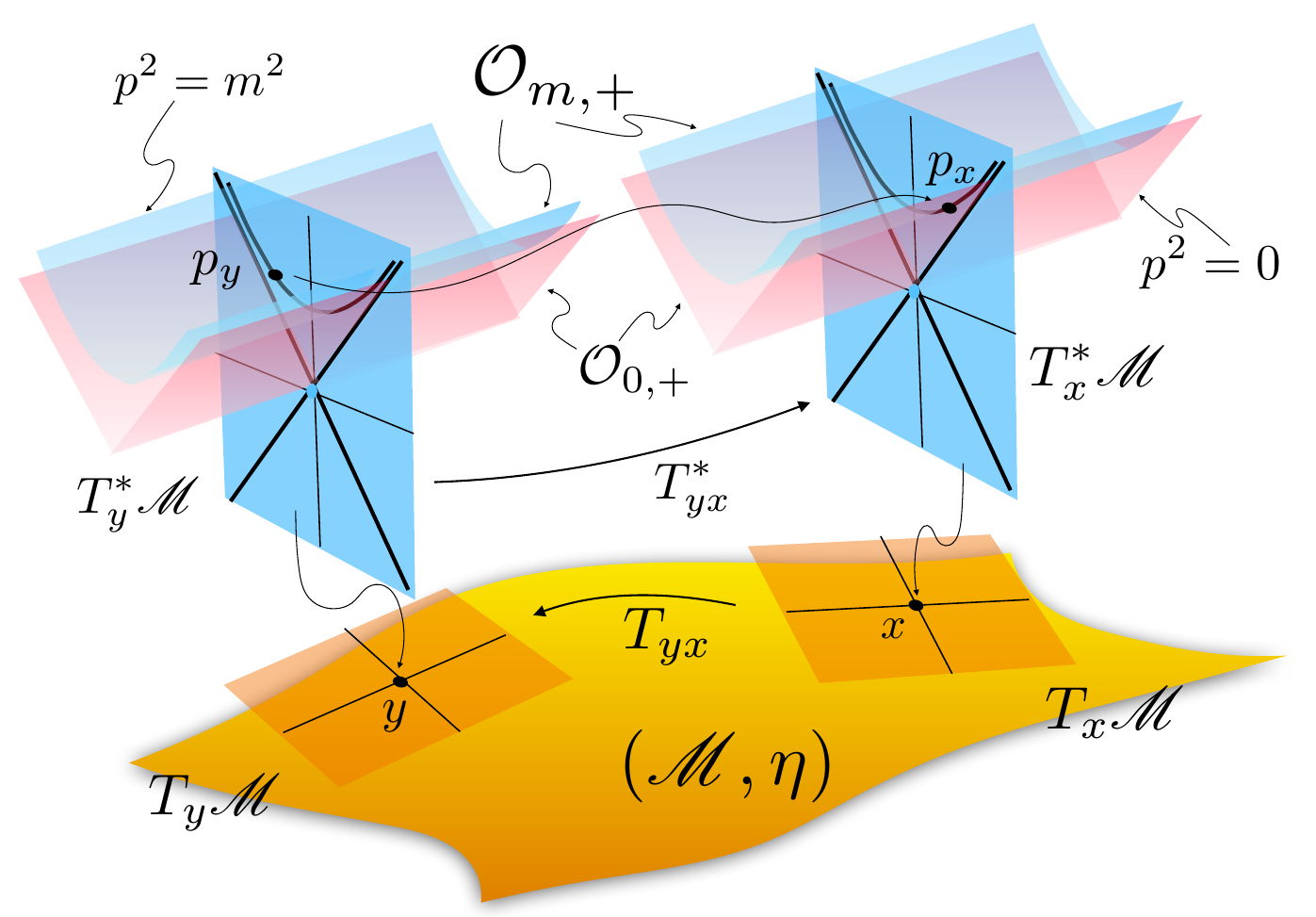}}
\vspace*{8pt}
\caption{A schematic illustration of the orbits of the Wigner groupoid of a space-time $(\mathscr{M}, \eta)$. Shadow blue: orbit $\mathcal{O}_{m,+}$; shadow purple: orbit $\mathcal{O}_{0,+}$. }\label{fig:orbits}
\end{figure}

\begin{example}[\textsc{Minkowski metric}]
Consider again Minkowski space-time $(\mathbb{M}^{1,3},\, \eta_0 )$.
In this case $\mathsf{Wigner}{(\mathscr{M},\,\eta)}$ reads
\be
\begin{split}
\mathsf{Wigner}{(\mathbb{M}^{1,3},\,  \eta_0 )} \,:=\, \bigl\{\, \left(p_y,\,T_{yx},\,p_x\right) \;\mid\;\; &p_x \in \mathbf{T}_x \mathbb{M}^{1,3},\, p_y \in \mathbf{T}_y\mathbb{M}^{1,3}\,, \\
&T_{yx}\in \mathrm{Hom}\left(\mathbf{T}_x \mathbb{M}^{1,3},\,\mathbf{T}_y \mathbb{M}^{1,3}\right) \;: \\
:\;\;&{T_{yx}}^* p_y \,=\, p_x \,, \;\; {T_{yx}}^*{ \eta_0 }_y \,=\, { \eta_0 }_x \,\bigr\} \,.
\end{split}
\ee
In this case the object space is the cotangent bundle $ \mathbf{T}^* \mathbb{M}^{1,3}$ which is a globally trivial bundle (since its base manifold is an affine space) which is actually an affine space itself, admitting a global chart with the associated system of coordinates $\left\{\, x^\mu,\, p_\mu \,\right\}_{\mu=0}^3$.
Again, the basis $\left\{\, e_I \,\right\}_{I=0}^3$ is global as well and is given by $\left\{\,\frac{\de}{\de x^\mu}\,\right\}_{\mu=0,...,3}$ modulo a local Lorentz transformation in the fixed tangent space $\mathbf{T}_x \mathbb{M}^{1,3}$.
In this case, the global trivialization
\be
\mathbf{T}\mathbb{M}^{1,3} \,=\, \mathbb{M}^{1,3} \times \mathbb{R}^4 \,,
\ee
provides the global decomposition of $\mathsf{Wigner}{(\mathbb{M}^{1,3},\, \eta_0 )}$
\be\label{eq:Minkowski_decomp}
\mathsf{Wigner}{(\mathbb{M}^{1,3},\, \eta_0 )} \,=\, \mathbb{M}^{1,3} \times \mathrm{O}(1,\,3) \times  \mathbf{T}^* \mathbb{M}^{1,3} \,,
\ee
given by:
$$
(p_y, T_{yx}, p _x) \mapsto (y, \Lambda, p_x) \, ,
$$
where $p_x \in   \mathbf{T}^*_x\mathbb{M}^{1,3}$,  $\Lambda \in O(1,3)$ is such that $T_{yx}\,  \frac{\partial}{\partial x^\mu}\mid_x = \Lambda_\mu^\nu \, \frac{\partial}{\partial x^\nu}\mid_y$ and $y \in \mathbb{M}^{1,3}$, showing that  $\mathsf{Wigner}{(\mathbb{M}^{1,3},\, \eta_0 )}$ is indeed a smooth manifold of dimension 18.  Note that $p_y$ is determined by $p_x$ and $\Lambda$, that is $(p_y)_\mu = \Lambda_\mu^\nu (p_x)_\nu$.     Moreover, from (\ref{eq:Minkowski_decomp}) and (\ref{eq:Poincare_decomp}) we get that the Wigner groupoid of Minkowski space-time is again an action groupoid, in this case the action groupoid corresponding to the natural action of Poincar\'e group on the phase space $ \mathbf{T}^* \mathbb{M}^{1,3}$ of Minkowski space-time:
$$
\mathsf{Wigner}{(\mathbb{M}^{1,3},\, \eta_0 )} = \mathcal{P} \times  \mathbf{T}^* \mathbb{M}^{1,3} \, .
$$
\end{example}

\begin{example}[\textsc{Schwarzschild metric}]
Consider again Schwarzschild space-time \((\mathbb{M}^{1,3}_{0},\,\eta_S)\).
Wigner's groupoid reads in this case:
\be
\begin{split}
\mathsf{Wigner}{(\mathbb{M}_0^{1,3},\, \eta_S)} \,:=\, \bigl\{\, \left(p_y,\,T_{yx},\,p_x\right) \;\mid\;\; &p_x \in \mathbf{T}_x \mathbb{M}_0^{1,3},\, p_y \in \mathbf{T}_y\mathbb{M}_0^{1,3}\,,\\
&T_{yx}\in \mathrm{Hom}\left(\mathbf{T}_x \mathbb{M}_0^{1,3},\,\mathbf{T}_y \mathbb{M}_0^{1,3}\right) \;: \\
:\;\;&{T_{yx}}^* p_y \,=\, p_x \,,\;\; {\eta_S}_y\left(T_{yx}(\,\cdot\,),\,T_{yx}(\,\cdot\,)\right) \,=\, {\eta_S}_x \,\bigr\} \,.
\end{split}
\ee

  The space of objects $ \mathbf{T}^* \mathbb{M}^{1,3}_0$ can be equipped with a global chart with the associated system of coordinates $\left\{\, U,\,V,\, \theta,\, \phi,\, p_I \,\right\}_{I=1,...,4}$, where $(U,\,V,\, \theta,\, \phi)$ are, again, the Kruskal coordinates, and  $p_I$ is the system of coordinates on the cotangent spaces associated to the choice of the basis of tetrad one-forms \eqref{Eq: tetrad one-forms Schwarzschild}.

   Then, for two charts \(U_k\) and \(U_j\) of $ \mathbf{T}^* \mathbb{M}^{1,3}_0$, an open set of the Wigner groupoid is given by
\[
\tilde{U}_{kj} \,=\, U_k \times U_{\mathrm{GL}_{kj}} \times U_j\,,
\]
where
\[
U_{\mathrm{GL}_{kj}} \,=\, \Bigl\{\, T_{yx}\colon \mathbf{T}_x U_j \to \mathbf{T}_y U_k \mid{T_{yx}}^* p_y \,=\, p_x \,,\;\, \eta_y\Bigl(T_{yx}(\cdot),T_{yx}(\cdot)\Bigr)=\eta_x \,\Bigr\}\,.
\]
Local coordinates on \(\tilde{U}_{kj}\) are provided by
\[
\tilde{\psi} \;:\; \tilde{U}_{kj}\to \mathbb{R}^8 \times \mathbb{R}^{16} \times \mathbb{R}^8\;:\; \xi \mapsto \tilde{\psi}(\xi) \,=\, \left(\,(x^\mu,\,{p_x}_I),\, \Lambda^I_J,\, (y^\mu,\, p_J)\,\right)\,,
\]
with the relation
\[
T_{yx}\, e_I(x) \,=\, \Lambda^J_I(x,\,y)\, e_J(y)\,,
\]
for $\Lambda^J_I(x,\,y)$ belonging to $O(1,\,3)$.
\end{example}
As for the Poincar\'e group, as it will be discussed below, the relevant part of Wigner groupoid for the classification of elementary particles is the connected component (as a topological space) containing the space of units, i.e., the space-time $\mathscr{M}$ itself as a submanifold of $\mathsf{Wigner}(\mathscr{M})$ by means of the canonical embedding $i \colon \mathscr{M} \to \mathsf{Wigner}(\mathscr{M})$, $i(x) = 1_x = (0_x, I_x, 0_x)$, with $0_x \in T^*_x\mathscr{M}$, the zero covector at $x$, and $I_x$ the identity map of the tangent space $T_x\mathscr{M}$, then the connected component of $\mathsf{Wigner}(\mathscr{M})$ containing $\mathscr{M}$ plays the role of the connected component of the identity in the theory of topological groups.

In the case of the Wigner groupoid $\mathsf{Wigner}{(\m,\,\eta)}$, the connected component containing its units, is the subset consisting of isometries $T_{yx}$ preserving both the orientation of $(\m,\,\eta)$ and its temporal orientation, namely, sending the future cone of $x$ into the future cone of $y$.   In what follows we will use the notation $\mathsf{Wigner}(\mathscr{M})$ to indicate the connected component of the units of the Wigner groupoid.


\section{The theory of projective representations of groupoids}
\label{Sec: Mackey's imprimitivity theorem and the theory of unitary representation of groupoids}

To state the main theorem of this work, we will need to extend the notion of projective representations of groups to the category of groupoids.
Such an extension amounts to introducing a new notion that replaces the Hilbert space associated with a given system by a field of Hilbert spaces \cite{Dixmier-Douady-ChampsContinus-1963, bos2011continuous}. Thus, the technical mathematical notion suitable for describing the theory of projective representations of groupoids is, at the same time, the notion that captures the physical setting determined by general space-times.

  In order to motivate all this, we will succinctly review the conceptual framework leading to Wigner's principle, and from there we will define the notion of a projective representation of a groupoid and construct an extension of the theory of induced representations of groups adapted to this context. The definitions introduced here are similar to, but not the same as, those introduced for instance in \cite{Pysiak-ImprimitivityGroupoids-2004, Pysiak-ImprimitivityGroupoids-2011}, and they follow the spirit of \cite{bos2011continuous}.

\subsection{Symmetries and projective representations of groupoids}\label{sec:projective}

We start by establishing the basic facts about symmetries of quantum systems. More precisely, let $\mathcal{H}_0$ be a complex separable Hilbert space associated with a given quantum system with inner product $\langle \cdot, \cdot \rangle_0$, and let $P(\mathcal{H}_0)$, the projective space of $\mathcal{H}_0$, denote its space of rays or pure states. The ray containing the nonzero vector $\psi \in \mathcal{H}_0$ will be denoted as $[\psi]$; thus $[\psi] = \{ \lambda \psi \;|\;\; \lambda \in \mathbb{C}\backslash \{0\}  \}$. Then, the transition probability defined by two rays $[\psi], [\phi] \in P(\mathcal{H}_0)$ is given by the real-valued function:
\begin{equation}\label{eq:transition}
\langle [\psi], [\phi] \rangle_0 = \frac{|\langle \psi, \phi \rangle_0|^2}{|| \psi ||_0^2 ||\phi ||_0^2} \,,
\end{equation}
where $\phi$ and $\psi$ are any representatives of $[\phi]$ and $[\psi]$.
An automorphism of $P(\mathcal{H}_0)$ is a bijective map $T \colon P(\mathcal{H}_0) \to P(\mathcal{H}_0)$ preserving the transition probabilities for any pair of rays:
$$
\langle T[\psi ], T[\phi ] \rangle_0 = \langle [\psi], [\phi] \rangle_0 \, , \quad \forall [\psi ], [\phi ] \in P(\mathcal{H}_0)\, .
$$
The family of all automorphisms of $P(\mathcal{H}_0)$ defines a group denoted as $\mathrm{Aut}(P(\mathcal{H}_0))$, and its elements describe the possible symmetries of the system described by $\mathcal{H}_0$ \cite{Bargmann-UnitaryRay-1954}.

  If $U$ is an (anti-)unitary operator on the Hilbert space $\mathcal{H}_0$, it induces an automorphism of $P(\mathcal{H}_0)$ by means of:
\begin{equation}\label{eq:TU}
T_U[\psi] := [U\psi] \, , \qquad \forall [\psi] \in P(\mathcal{H}_0) \, .
\end{equation}
If we denote by $\widetilde{U}(\mathcal{H}_0)$ the group of (anti-)unitary operators on $\mathcal{H}_0$, then Wigner's theorem states that we have the following short exact sequence of groups \cite{simms1968lie}:
\begin{equation}\label{eq:wignerthm}
1 \to U(1) \to \widetilde{U}(\mathcal{H}_0) \to \mathrm{Aut} (P(\mathcal{H}_0)) \to 1 \, ,
\end{equation}
which means that any automorphism $T$ of $P(\mathcal{H}_0)$ is of the form $T_U$ for some (anti-)unitary operator $U$. The connected component of the group $\mathrm{Aut} (P(\mathcal{H}_0)) $ will be denoted by $PU(\mathcal{H}_0)$ and is the projection of the connected component of the group $\widetilde{U}(\mathcal{H}_0)$, i.e., the group $U(\mathcal{H}_0)$ of unitary maps on $\mathcal{H}_0$.
Hence, a Lie group $G$ will be a symmetry for the quantum system with associated Hilbert space $\mathcal{H}_0$ if there is a continuous\footnote{Continuity here means continuity with respect to the topology induced on the group $\mathrm{Aut}(P(\mathcal{H})_0)$ by the strong topology on the group $\widetilde{U}(\mathcal{H}_0)$.} map $T \colon G \to \mathrm{Aut}(P(\mathcal{H}_0))$ such that:
\begin{equation}\label{eq:rep_group}
T(gg') = T(g) T(g') \, , \qquad T(e) = \mathrm{id}_{\mathcal{H}_0} \, ,\;\;\; \forall \,\,g,g' \in G \, .
\end{equation}
Such a map will be called a projective representation of the Lie group $G$ and, in categorical language, it is a functor:
\begin{equation}\label{eq:functor}
T\colon G \to \mathsf{Proj} \, ,
\end{equation}
from the category defined by the group $G$ itself to the category $\mathsf{Proj}$ whose objects are projective spaces and whose morphisms are maps preserving transition probabilities, such that $T$ is continuous in the sense defined above.

  If the Lie group $G$ is connected, because of continuity, a projective representation of $G$ will be given by a continuous homomorphism of groups:
\begin{equation}\label{eq:proj_connected}
T \colon G \to PU(\mathcal{H}_0) \, .
\end{equation}
In particular, if $G$ is a group of symmetries of a given space-time $\mathscr{M}$, it will be natural to consider that such a group is implemented as a group of symmetries of any quantum system defined on such a space-time.
Thus, if the space-time is Minkowski space-time, $\mathscr{M} = \mathbb{M}^{1,3}$, we will ask the Poincar\'e group to be represented projectively on the Hilbert space of the system \cite{Bargmann-UnitaryRay-1954}.

  Projective representations of a connected Lie group $G$ are in one-to-one correspondence with a family of unitary representations of a universal extension $\overline{G}$ of $G$ called the \textit{projective covering group} of $G$ \cite[Theorem 3.2]{Carinena-Santander-ProjectiveUnitary-1975}. The fundamental idea is that the obstructions to lifting a projective representation to a unitary one are cohomological in nature and can be understood at the level of the Lie algebra $\mathfrak{g}$ of $G$.

The construction of \cite{Carinena-Santander-ProjectiveUnitary-1975} begins with the computation of the second Lie algebra cohomology group, $H^2(\mathfrak{g}, \mathbb{R})$, of the Lie algebra of the Lie group $G$.
This group classifies the non-trivial central extensions of $\mathfrak{g}$.
If we choose a family $\omega_i$ of representatives of the classes $[\omega_i]$, $i = 1, \ldots, r$, $r = \dim H^2(\mathfrak{g}, \mathbb{R})$, defining a basis of $H^2(\mathfrak{g}, \mathbb{R})$, the Lie algebra of the projective covering group, denoted by $\overline{\mathfrak{g}}$, is a central extension of $\mathfrak{g}$ given by:
\be
\overline{\mathfrak{g}} = \mathfrak{g} \oplus \mathbb{R}^{\dim H^2(\mathfrak{g}, \mathbb{R})} \, ,
\ee
where the Lie bracket on $\overline{\mathfrak{g}}$ is a deformation of the bracket on $\mathfrak{g}$, determined by the 2-cocycles representing the cohomology classes.
For any pair of elements $\overline{X} = (X, a), \overline{Y} = (Y,b) \in \overline{\mathfrak{g}}$, the new bracket is given by:
\be
[\overline{X}, \overline{Y}]_{\overline{\mathfrak{g}}} = ([X, Y]_{\mathfrak{g}}, \omega_1(X, Y), \ldots, \omega_r(X, Y)) \, .
\ee
The projective covering group $\overline{G}$ is then the connected and simply connected Lie group whose Lie algebra is $\overline{\mathfrak{g}}$ (see Sect. \ref{sec:massless} for more details and the application of this theorem to the computation of the projective representations of the Euclidean group $E(2)$).

It is worth stressing that if $H^2(\mathfrak{g}, \mathbb{R}) = 0$, $\mathfrak{g}$ being the Lie algebra of $G$, the projective covering group coincides with the universal covering group of $G$.
This is indeed the situation for the Poincar\'e group $\mathcal{P}$, and this is the reason why we classify elementary particles as irreducible unitary representations of the universal covering $\widetilde{\mathcal{P}}$ of the Poincar\'e group in \cref{sec:wigner_classification}. If we denote by $p \colon \overline{G} \to G$ the covering homomorphism, then we have the following commutative diagram:

\begin{figure}[h!]\begin{center}
\begin{tikzpicture}
\draw [->] (0.4,0) -- (1.4,0);
\draw [->] (2.6,0) -- (3.6,0);
\draw [->] (4.4,0) -- (5.6,0);
\draw [->] (6.4,0) -- (7.6,0);
\draw [->] (2.6,-1.5) -- (3.4,-1.5);
\draw [->] (4.6,-1.5) -- (5.2,-1.5);
\draw [->] (1.9,-0.3) -- (1.9,-1.2);
\draw [->] (5.9,-0.3) -- (5.9,-1.2);
\draw [->] (4,-0.3) -- (4,-1.2);
\draw [->] (0.4,-1.5) -- (1.4,-1.5);
\draw [->] (6.8,-1.5) -- (7.6,-1.5);
\node  at (0,0) {$1$};
\node  at (2,0) {$\ker p$};
\node(tilGam) at (4,0) {$\overline{G}$};
\node(P)  at (6,0) {$G$};
\node  at (8,0) {$1$};
\node  at (0,-1.5) {$1$};
\node  at (2,-1.5) {$U(1)$};
\node(Omega2)  at (4,-1.5) {$U(\mathcal{H}_0)$};
\node(Sigma)  at (6,-1.5) {$PU(\mathcal{H}_0)$};
\node  at (8,-1.5) {$1$};
\node  at (3,0.2) {$i$};
\node  at (5,0.2) {$p$};
\node  at (5,-1.3) {$\pi$};
\node  at (6.2,-0.7) {$T$};
\node  at (4.3,-0.7) {$R$};
\end{tikzpicture}
\caption{Diagram describing the relation between projective representations of a connected Lie group and the unitary representations of its projective covering group $\overline{G}$.}\label{fig:extension}
\end{center}
\end{figure}

  Each projective representation $T\colon G \to PU(\mathcal{H}_0)$ can be lifted to a unitary representation $R\colon \overline{G} \to U(\mathcal{H}_0)$. Conversely, each unitary representation $R$ of $\overline{G}$ mapping the kernel of $p$ into $U(1)$ induces a projective representation of $G$ such that:
\be
T\circ p = \pi \circ R \, .
\ee

\bigskip

When the given space-time $\mathscr{M}$ is non-trivial\footnote{As was discussed in the introduction, when $\mathscr{M}$ is not a homogeneous space of a relativistic kinematical group.}, it is not obvious how to associate a Hilbert space with a system living in such a space-time. As was stated in the introduction, we do not want to move in the direction of quantum field theories. Instead, we will modify the notion of symmetry, assuming that the symmetry of the system is captured by the Wigner groupoid of the given space-time $\mathscr{M}$. Then, in order to introduce the appropriate notion of representation of the Wigner groupoid, we have to assume that instead of having a fixed Hilbert space $\mathcal{H}_0$ associated with the overall system, we have a family of Hilbert spaces $\mathcal{H}_{(x,p)}$ associated with each event $\xi = (x,p) \in T^\star\mathscr{M}$. In fact, we can assume that, in this particular instance, the Hilbert spaces $\mathcal{H}_{(x,p)}$ do not depend on $p$ and depend only on the space-time event $x$, that is $\mathcal{H}_{(x,p)} = \mathcal{H}_x$. This idea is akin to the notion of associating algebras of observables to open sets in space-time, instrumental in AQFT.

  Thus, we will posit that there is a Hilbert space associated to any event $x \in \mathscr{M}$, $x \mapsto \mathcal{H}_x$. Moreover, there will be a family $\mathcal{S}$ of functions $\psi \colon \mathscr{M} \to \prod_{x\in \m} \mathcal{H}_x$, $\psi (x) \in \mathcal{H}_x$, satisfying a few natural consistency conditions:
\begin{enumerate}
\item $\{ \psi (x) \;\mid\;\; \psi \in \mathcal{S} \} = \mathcal{H}_x$, for all $x \in \m$;
\item For all $\psi,\phi \in \mathcal{S}$, the evaluation map $x \mapsto  \langle \psi (x), \phi (x) \rangle_x$, where $\langle\cdot, \cdot \rangle_x$ denotes the inner product in the Hilbert space $\mathcal{H}_x$, is continuous.
\end{enumerate}
A family of Hilbert spaces satisfying the previous conditions (together with a technical uniform closedness condition which will not be relevant in the current context) is called a continuous field of Hilbert spaces (see \cite[Def. 2.20, (iii)]{Bos-ContinuousRepresentations-2006}, \cite[Def. 1, CCiv]{Dixmier-Douady-ChampsContinus-1963}) and represents a natural extension of the fundamental quantum mechanical principle that assigns a Hilbert space to a given quantum system when such a system has an associated space-time of events $\mathscr{M}$.

  It is natural to assign to any continuous field of Hilbert spaces $(\{ \mathcal{H}_x\}_{x \in \mathscr{M}}, \mathcal{S})$ a topological bundle $\tau \colon \mathcal{H} \to \mathscr{M}$ such that the family of maps $\mathcal{S}$ can be identified with the continuous sections of $\tau$. Indeed, we define the total space of the bundle as $\mathcal{H} = \sqcup_{x\in \mathscr{M}} \mathcal{H}_x$, with the canonical map $\tau (\psi_x) = x$, and the topology given by the family of open sets:
$$
U(\epsilon, \psi ,V) :=  \{  \zeta \in  \mathcal{H} \mid || \zeta - \psi(\tau(\zeta))||  < \epsilon \mathrm{\, \,\,and\, \,}  \tau(\zeta) \in V \} \, ,
$$
for any $\epsilon > 0$, $V \subset \mathscr{M}$ an open subset, and $\psi \in \mathcal{S}$. Then it can be proved that $\mathcal{S} = \Gamma^0 (\mathcal{H})$  (see \cite[Lemma 1.5]{Bos-ContinuousRepresentations-2006}), i.e., $\mathcal{S}$ is the set of continuous sections of the bundle $\tau$.
We say that a continuous field of Hilbert spaces is locally trivial if the associated bundle $\tau \colon \mathcal{H} \to \mathscr{M}$ is locally trivial, an assumption that will be made in what follows. The bundle $\tau\colon \mathcal{H} \to \mathscr{M}$ is a Hermitian bundle of Hilbert spaces, that is, it carries (by construction) a continuous family of inner products along its fibres, given by $x \mapsto \langle\cdot, \cdot \rangle_x$.

\begin{definition}\label{def:space-timefields}
Given a space-time $\mathscr{M}$, we will say that a locally trivial continuous field of Hilbert spaces $(\{ \mathcal{H}_x\}_{x \in \mathscr{M}}, \mathcal{S})$ is a space-time adapted continuous field over $\mathscr{M}$, and we will identify it with its associated Hermitian bundle $\tau \colon \mathcal{H} \to \mathscr{M}$.
\end{definition}

  Given an adapted continuous field of Hilbert spaces with bundle $\tau \colon \mathcal{H} \to \mathscr{M}$, there is a canonical bundle of projective spaces associated with it and denoted by $p_\tau: P(\mathcal{H}) \to \mathscr{M}$, whose fibres are the projective spaces $P(\mathcal{H}_x)$ and whose topology is the quotient topology of $\mathcal{H}$, with respect to the canonical projection $\rho \colon \mathcal{H} \to P(\mathcal{H})$, $\rho (\psi_x) = [\psi_x]$, for all $\psi_x \in p^{-1}(x) = \mathcal{H}_x$.

If the symmetry of a quantum system is carried by a groupoid instead of a group, we need to transport the previous discussion and definitions to the category of groupoids, giving rise to the notion of a projective representation of a Lie groupoid.

Thus, if we had a groupoid $\mathcal{G} \rightrightarrows \mathscr{M}$ describing the symmetry of a certain quantum system over the space-time $\mathscr{M}$ determined by an adapted field of Hilbert spaces $\mathcal{H}$, then for any morphism $\alpha \colon x \to y$ the transition probabilities perceived at the point $x$ should be the same as the transition probabilities perceived at the point $y$ after the action of the morphism $\alpha\in \mathcal{G}$. In other words, associated with the morphism $\alpha \colon x \to y$, there must be a map $\Phi(\alpha) \colon P(\mathcal{H}_x) \to P(\mathcal{H}_y)$, where $\mathcal{H}_x$ is the Hilbert space associated with the system at the point $x$ and similarly for $y$, such that:
$$
\langle \Phi_\alpha([\psi_x]), \Phi_\alpha([\phi_x]) \rangle_y := \frac{ |\langle \Phi(\psi_x) , \Phi(\phi_x) \rangle_y |^2}{||\Phi(\psi_x)||_y^2 ||\Phi(\phi_x)||_y^2} = \frac{ |\langle \psi_x , \phi_x \rangle_x |^2}{||\psi_x||_x^2 ||\phi_x||_x^2} =: \langle [\psi_x], [\phi_x] \rangle_x\, ,
$$
for every $[\psi_x], [\phi_x] \in P(\mathcal{H}_x)$.

  Because of Wigner's theorem, such a map $\Phi(\alpha)$ is of the form $T_{U(\alpha)}$ (see \eqref{eq:TU}), where $U(\alpha) \colon \mathcal{H}_x \to \mathcal{H}_y$ is a (anti-)unitary map. The consistency of the composition of morphisms $\alpha$, $\beta$ in the groupoid and their associated maps $\Phi(\alpha)$, $\Phi (\beta)$, requires that, as in \eqref{eq:rep_group}:
\begin{equation}\label{eq:Ralphabeta}
\Phi(\alpha \circ \beta ) = \Phi(\alpha) \circ \Phi(\beta) \, ,
\end{equation}
for all composable pairs of morphisms $(\alpha, \beta) \in \mathcal{G}^{(2)}$,
and the map associated to the unit $1_x \colon x \to x$ in the groupoid $\mathcal{G}$ must be the identity, that is:
\begin{equation}\label{eq:Runits}
\Phi(1_x) = \mathrm{id\,}_{\mathcal{H}_x} \, , \qquad \forall x \in \mathscr{M} \,.
\end{equation}

However, in general, the groupoids that will be of interest to us, like the Wigner groupoid, have spaces of units which are different from the space-time $\mathscr{M}$ we are interested in (in the case of the Wigner groupoid, its space of units is $\mathbf{T}^*\mathscr{M}$). Thus, we will provide the definition of projective representations of Lie groupoids $\mathcal{G} \rightrightarrows \Omega$ with a general space of units $\Omega$.

\begin{definition}[\textsc{Projective Representation of a Lie Groupoid}]\label{def:projective}
Let \(\mathcal{G} \rightrightarrows \Omega\) be a Lie groupoid and let $\tau \colon \mathcal{H} \to \Omega$ be a bundle of Hilbert spaces describing a locally trivial continuous field of Hilbert spaces over the manifold $\Omega$. A continuous projective representation (or just a projective representation if there is no risk of confusion) of the groupoid $\mathcal{G}$ is a family of bijective maps $\Phi (\alpha) \colon P(\mathcal{H}_x) \to P(\mathcal{H}_y)$, for every morphism $\alpha \colon x \to y \in \mathcal{G}$, $x,y \in \Omega$, satisfying conditions \eqref{eq:Ralphabeta}-\eqref{eq:Runits}, and such that it is continuous, that is, such that the map:
\begin{eqnarray}
&& \mathrm{ev\,} \colon \mathcal{G}\times_\Omega P(\mathcal{H}) = \{ (\alpha, [\psi]) \in \mathcal{G} \times P(\mathcal{H}) \mid s(\alpha) =  p[\psi] \}  \to P(\mathcal{H}) \nonumber \\ &&\qquad (\alpha, [\psi]) \mapsto \Phi(\alpha)[\psi] \, , \label{eq:continuity}
\end{eqnarray}
is continuous.
We will denote such a representation as $(\Phi, \tau)$ (or just $\Phi$ if there is no risk of confusion) and $\Phi_\alpha := \Phi (\alpha)$ for all $\alpha\in \mathcal{G}$. We will also say that the bundle $\tau \colon \mathcal{H}\to \Omega$ supports the representation $\Phi$.
\end{definition}

\begin{remark}
The previous definition of a projective representation of a groupoid $\mathcal{G}$ extends in a natural way the notion of a projective representation of a Lie group stated before (see \eqref{eq:proj_connected}). In fact, in categorical language, a projective representation of a groupoid is just a functor $\Phi \colon \mathcal{G} \to \mathsf{Proj}$, satisfying the continuity condition \eqref{eq:continuity}, which is exactly the same categorical definition of a projective representation of a Lie group $G$, \eqref{eq:functor}.
\end{remark}

\begin{remark}
The definition of a continuous projective representation presented here is valid in a straightforward way in the category of topological groupoids (more precisely, locally compact topological groupoids to avoid difficulties with the existence of appropriate measures); however, it is established just for Lie groupoids because those are the groupoids we are dealing with in this paper.
\end{remark}

\begin{remark}
We have just stated the definition of projective representations of Lie groupoids, but the notion of continuous unitary representations of Lie groupoids is defined in an obvious way, that is, as functors from the given groupoid to the category $\mathsf{Hilb}$ of Hilbert spaces and bounded operators, satisfying a natural continuity condition like \eqref{eq:continuity}.
\end{remark}

\begin{remark}
It can be shown that the continuity condition \eqref{eq:continuity} is equivalent to other continuity conditions like strong or weak continuity for unitary representations \cite[Ch. 2]{Bos-ContinuousRepresentations-2006}.
\end{remark}

  A projective subbundle of the bundle of projective spaces $p_\tau \colon P(\mathcal{H}) \to \Omega$ associated to the Hermitian bundle $\tau \colon \mathcal{H} \to \Omega$ is the projection of a subbundle $E \subset \mathcal{H}$ of closed Hilbert subspaces $E_x \subset \mathcal{H}_x$, $x \in \Omega$.

\begin{definition}[\textsc{Irreducible Representation}]
A projective representation \((\Phi,\, \tau)\) of the Lie groupoid $\mathcal{G}$ is said to be irreducible if it admits no nontrivial invariant projective subbundles.
That is, the only subbundles ${p_\tau}_W\colon  P(W) \to \Omega$ of $p_\tau \colon P(\mathcal{H}) \to \Omega$ such that \(\Phi_\alpha(P(W_{s(\alpha)}) \subset P(W_{t(\alpha)})\) for all \(\alpha \in \mathcal{G}\) are either \( P(W) = \{0\}\) or \( P(W) = P(\mathcal{H})\).
\end{definition}

\begin{definition}[\textsc{Equivalence of Representations}]\label{def:equivalence}
Given two projective representations \((\Phi,\, \tau)\) and \((\Phi',\, \tau')\) of the Lie groupoid $\mathcal{G}$, we will say that they are equivalent if there is an isomorphism of the associated projective bundles $\chi \colon P(\mathcal{H}) \to P(\mathcal{H}')$, such that:
\begin{equation}\label{eq:unitary_equiv}
\Phi(\alpha)  \circ \chi (s(\alpha)) = \chi (t(\alpha)) \circ \Phi (\alpha) \, , \qquad \forall \alpha \in \mathcal{G} \, .
\end{equation}
\end{definition}

  In the previous definition, an isomorphism $\chi  \colon P( \mathcal{H}) \to P(\mathcal{H}')$ is a continuous map such that $p' \circ \chi = p$, and for each $x \in \Omega$, the restriction $\chi\mid_{P(\mathcal{H}_x)} \colon P(\mathcal{H}_x) \to P(\mathcal{H}'_x)$ is an automorphism of projective spaces.

\begin{remark}
Because of Wigner's theorem, this notion is equivalent to establishing that two projective representations \((\Phi,\, \tau)\) and \((\Phi',\, \tau')\) are equivalent if there exists an isomorphism $\tilde{\chi} \colon \mathcal{H} \to \mathcal{H}'$ of their corresponding Hermitian bundles.
\end{remark}

\begin{remark}\label{rem:irreducible}
The notion of equivalence of projective representations introduced here amounts to saying that there is an invertible natural transformation $\chi \colon \Phi \to \Phi'$ among the functors determining the projective representations.
\end{remark}

  If $\Phi_1$, $\Phi_2$ are projective representations of the Lie groupoids $\mathcal{G}_1\rightrightarrows \Omega_1$ and $\mathcal{G}_2 \rightrightarrows \Omega_2$ supported on the Hermitian bundles $\tau_1 \colon \mathcal{H}_1 \to \Omega_1$ and $\tau_2\colon \mathcal{H}_2\to \Omega_2$ respectively, we can form the direct sum representation $\Phi: = \Phi_1\oplus \Phi_2$ of the disjoint union of the groupoids $\mathcal{G}_1$ and $\mathcal{G}_2$, $\mathcal{G}_1 \sqcup \mathcal{G}_2 \rightrightarrows \Omega_1\sqcup \Omega_2$, given by the assignment $\Phi(\alpha_a) = \Phi_a (\alpha_a) \colon P(\mathcal{H}_{x_a}) \to P(\mathcal{H}_{y_a})$, $\alpha_a \colon x_a \to y_a \in \mathcal{G}_a$, $a = 1,2$ (\cite[Prop. 9.5]{Ibort-Rodriguez-IntroGroupoids-2019}).
When dealing with algebraic groupoids, projective representations of groupoids decompose into the direct sum, in the previous sense of projective representations, of the restriction of the groupoid $\mathcal{G}$ to its orbits (recall Eq. \eqref{eq:structure}); that is (see also, \cite[Prop. 9.6]{Ibort-Rodriguez-IntroGroupoids-2019}):
\begin{equation}\label{eq:decomposition_rep}
\Phi = \bigoplus_{O \in \Omega/\mathcal{G}} \Phi_O \, ,
\end{equation}
where $\Phi_O \colon \mathcal{G}_O \to \mathsf{Proj}$ is the restriction of the projective representation $\Phi$ to the subgroupoid $\mathcal{G}_O \subset \mathcal{G}$ obtained by restricting $\mathcal{G}$ to the orbit $O$.
In this sense, when dealing with non-transitive Lie groupoids $\mathcal{G}$, we will say that an irreducible projective representation of such a groupoid is an irreducible projective representation of any of its transitive subgroupoids $\mathcal{G}_O$.


\subsection{Induced projective representations of Lie groupoids}\label{sec:inducing}

Now we can state precisely the relation between projective representations of a transitive groupoid \(\mathcal{G} \rightrightarrows \Omega\) over the manifold $\Omega$ and the projective representations of its isotropy group $G$. We will assume that both $\Omega$ and the isotropy group $G$ are connected as topological spaces.

  We choose an arbitrary point $x_0 \in \Omega$. Because all isotropy groups are isomorphic, we identify $G$ with the isotropy group at $x_0$, which is a connected Lie group by assumption:
\be
G := G_{x_0} =  \{\, \alpha \in \mathcal{G} \;:\;\; s(\alpha) \,=\, t(\alpha) \,=\, x_0 \,\} \, .
\ee

The correspondence between projective representations of the groupoid $\mathcal{G}$ and those of the isotropy group $G$, as for the classical Mackey theorem, is trivial in one direction; namely, the restriction $\Phi\mid_{G_0}$ of any projective representation $\Phi$ of the groupoid $\mathcal{G}$ supported in the projective bundle $P(\mathcal{H})$ defines a projective representation of $G$. Thus, if we denote by $\mathrm{Proj}(\mathcal{G})$ and $\mathrm{Proj} (G)$ the spaces of equivalence classes of projective representations of $\mathcal{G}$ and $G$, respectively, we have a natural map $r \colon \mathrm{Proj}(\mathcal{G}) \to \mathrm{Proj}(G)$ by restriction.  Note that if $P(\mathcal{H})$ is irreducible under the groupoid action, then $P(\mathcal{H}_{x_0})$ is irreducible under the action of $G$. 

However, the inverse construction is non-trivial.   The main idea is to construct a bundle $E_T$ by ``twisting'' the groupoid with the representation of the isotropy group, similar to how one constructs an associated bundle in gauge theory.  Thus, consider a projective representation $T \colon G \to PU(\mathcal{H}_0)$, supported on the Hilbert space $\mathcal{H}_0$. Let $\mathcal{G}_{x_0}$ denote the embedded submanifold $s^{-1}(x_0) = \{ \alpha \colon x_0 \to y\in \mathcal{G} \mid y \in \Omega \}$ (recall \cref{sec:groupoids_symmetry}).
Then, the set:
\begin{equation}\label{eq:associated}
E_T\,:=\,\mathcal{G}_{x_0} \times P(\mathcal{H}_0) / G \, ,
\end{equation}
where $G$ acts freely on the right on $\mathcal{G}_{x_0} \times P(\mathcal{H}_0)$ as:
$$
(\alpha,\, [\phi]) \cdot \gamma = (\alpha \circ \gamma,\, T(\gamma)^{-1} [\phi])\,, \qquad \forall  \alpha \in \mathcal{G}_{x_0},\, [\phi] \in P(\mathcal{H}_0),\, \gamma \in G \, ,
$$
is a projective bundle over $\Omega$. Indeed, the points in $E_T$ are equivalence classes:
\begin{equation}\label{eq:projectivefibre}
[\alpha,\,[\phi]] \,:=\, \left\{\, 
(\alpha \circ \gamma,\, T(\gamma)^{-1} [\phi]) \mid  \gamma \in G \,\right\} \, ,
\end{equation}
and $E_T$ inherits the quotient topology from the product topology on $\mathcal{G}_{x_0} \times P(\mathcal{H}_0)$. Then, the canonical projection map:
\be
\pi_T \colon E_T \to \Omega \, , \qquad   [\alpha,\, [\phi]] \mapsto \pi_T ([\alpha,\, [\phi]]) \,=\, t(\alpha) \,,
\ee
is a continuous fibre bundle projection.

  Given a morphism $\alpha\colon x_0 \to x$, the fibre $\pi_{T}^{-1}(x) \,=:\, (E_T)_x$, i.e., those classes $[\alpha, [\psi]]$ such that $t(\alpha) = x$, can be identified with the projective space $P(\mathcal{H}_0)$ by means of the homeomorphism $i_\alpha\colon P(\mathcal{H}_{0}) \to (E_T)_x$ given by:
\begin{equation}\label{eq:embedding}
i_\alpha ([\phi]) = [\alpha, [\phi]] \,.
\end{equation}
Note that if $\alpha' \colon x_0 \to x$ is another morphism relating $x_0$ and $x$, then
\begin{equation}\label{eq:embedding_prime}
i_\alpha' = i_\alpha \circ T(\gamma)\, , \qquad \mathrm{with\,\,\,} \gamma = \alpha^{-1} \circ \alpha' \in G \, .
\end{equation}
The embedding \eqref{eq:embedding} allows us to define transition probability functions along the fibres of $E_T$:
 $$
 \langle \cdot, \cdot \rangle_x \colon (E_T)_x \to \mathbb{R} \, , \qquad \forall x \in \Omega \, ,
$$
by means of:
\begin{equation}\label{eq:transition_fibres}
\langle i_\alpha(\psi), i_\alpha (\psi) \rangle_x \,=\, \langle [\alpha, [\psi]], [\alpha, [\psi]] \rangle_x = \langle [\psi], [\phi] \rangle_0 \, .
\end{equation}
Let us point out that because of \eqref{eq:embedding_prime}, the expression on the r.h.s. of \eqref{eq:transition_fibres} is well defined.

It is easy to show that the bundle $\pi_T \colon E_T \to \Omega$ is locally trivial by choosing a local section $\sigma \colon U \subset \Omega \to \mathcal{G}_{x_0}$ of the canonical projection $\nu \colon \mathcal{G}_{x_0} \to \mathcal{G}_{x_0}/G \cong \Omega$. Then, we define:
$$
\varphi \colon \pi_T^{-1}(U) \to U \times P(\mathcal{H}_{x_0}) \, , \qquad \varphi (\alpha) = (t(\alpha), i_\alpha^{-1}(\sigma (t(\alpha)))) \, .
$$
It is worth pointing out that $\nu\colon  \mathcal{G}_{x_0} \to \Omega$ is a principal bundle over $\Omega$ with structure group $G$, and the space $E_T$ defined in (\ref{eq:associated}) is just the associated bundle to $\nu$ defined by the projective representation $T$.
 
\begin{remark}  In general, given a projective bundle $\pi_E \colon E \to \Omega$, it is not true that there exists a bundle of Hilbert spaces $\pi_\mathcal{H} \colon \mathcal{H} \to \Omega$ such that $E$ is the projectivization $P(\mathcal{H})$ of the bundle $\mathcal{H}$. The obstruction for the existence of such an unfolding of the projective bundle $E$ is determined by a cohomology class $\delta_E \in H^3(\Omega, \mathbb{Z})$, called the Dixmier-Douady class of $E$ \cite{Dixmier-Douady-ChampsContinus-1963}. However, we will show that the projective bundle $E_T$ associated with the projective representation $T$ will be the projectivization of a Hilbert bundle by constructing explicitly the bundle $\pi_\mathcal{H} \colon \mathcal{H} \to \Omega$ such that $P(\mathcal{H}) = E_T$. 
This will be the content of the main theorem in the coming section.
\end{remark}


\subsection{The classification of projective representations of groupoids}\label{sec:projective_rep_groupoids}

  Now, we can state the main theorem of this section.

\begin{theorem} \label{Thm: Mackey theorem transitive groupoids}
Let \(\mathcal{G} \rightrightarrows \Omega\) be a transitive Lie groupoid such that its isotropy group $G$ is connected. Given a projective representation $T \colon G \to PU(\mathcal{H}_0) \in \mathrm{Proj}(G)$, there exists a projective representation $\Phi \colon \mathcal{G} \to \mathsf{Proj} \in \mathrm{Proj}(\mathcal{G})$, such that $r(\Phi) = T$ or, equivalently, the Dixmier-Douady class $\delta (E_T)$ vanishes. It will be said that the projective representation of the groupoid associated with a given projective representation $T$ is induced from $T$ and denoted $\Phi = \mathrm{Ind}_G^\mathcal{G}(T)$ (or just $ \mathrm{Ind\,}(T)$ if there is no risk of confusion).
In particular, if the projective representation $T$ of $G$ is irreducible, its induced representation is irreducible too.
\end{theorem}

\begin{proof}
  The bundle $\pi_T \colon E_T \to \Omega$, Eq. (\ref{eq:associated}), associated with the projective representation $T$ of $G$ carries a representation of $\mathcal{G}$ given by:
\begin{align}
\mathscr{M} &\ni x \mapsto (E_T)_x = \pi_T^{-1}(x)\,, \\
\mathcal{G} &\ni \beta \colon x \to y \mapsto  \Phi_\beta \;:\;\; (E_T)_x \to (E_T)_y \,, 
\end{align}
where $x \,=\, s(\beta)$, $y \,=\, t(\beta)$, and
\begin{equation}\label{eq:ind_rep}
\Phi_\beta([\alpha,\, [\phi]]) \,=\, [\beta \circ \alpha,\, [\phi]] \, , \qquad \forall [\alpha, [\psi]] \in (E_T)_x \, ,
\end{equation}
as the following chain of equalities shows:
\be
\Phi_{\delta \circ \beta} ([\alpha,\, [\phi]]) \,=\,[ (\delta \circ \beta) \circ \alpha,\, [\phi] ]\,=\, [\delta \circ (\beta \circ \alpha),\, [\phi]] \,=\, \Phi_\delta \Phi_\beta [\alpha,\, [\phi] ]\, .
\ee
 
  Note that the representation $\Phi$ preserves the transition probabilities because of \eqref{eq:transition_fibres} and the chain of equalities below:
\begin{eqnarray*}
\langle \Phi_\beta([\alpha,\,[\phi]]),\, \Phi_\beta([\alpha,\, [\psi]])\rangle_y  &=& \langle [\beta \circ \alpha,\,[\phi]],\, [\beta \circ \alpha,\, [\psi]])\rangle_y \\ &=& \langle i_{\beta \circ \alpha}([\phi]),\, i_{\beta \circ \alpha} ([\psi])\rangle_y = \langle [\phi], [\psi]\rangle_0 \\
&=& \langle [\alpha,\, [\phi]],\, [\alpha,\, [\psi]]\rangle_x \, .
\end{eqnarray*}

  It is immediate to check the continuity of the representation $\Phi$ (see \eqref{eq:continuity}, \cref{def:projective}). Indeed, formula \eqref{eq:ind_rep} shows that $\Phi_\beta$ is continuous because the composition $\circ$ is smooth and $E_T$ inherits the quotient topology from the product $\mathcal{G}_{x_0} \times \mathcal{H}_0$.

  Once the representation $\Phi$ of $\mathcal{G}$ with support on the projective bundle $E_T$ has been constructed, it remains to be seen that there exists an adapted field of Hilbert spaces $\mathcal{H}$ such that $E_T$ is its projectivization, that is, $E _T\cong P(\mathcal{H})$. We will provide an explicit construction of the field of Hilbert spaces $\mathcal{H}$ by using the projective covering group $\overline{G}$ (recall \cref{fig:extension} and \cite{Carinena-Santander-ProjectiveUnitary-1975}). Given the representation $T\colon G \to PU(\mathcal{H}_0)$, there exists a unitary representation $R \colon \overline{G} \to U(\mathcal{H}_0)$ such that $R(A) \subset U(1)$ and which passes to the quotient with respect to the epimorphism $p \colon \overline{G}  \to G$, giving the projective representation $T$ of $G$. Then, we construct a bundle of Hilbert spaces $\mathcal{H}$ defined as:
\begin{equation}
\mathcal{H} = \mathcal{G}_{x_0} \times \mathcal{H}_0 / \overline{G} \, ,
\end{equation}
where $\overline{G}$ acts on the right on $\mathcal{G}_{x_0} \times \mathcal{H}_0$ by means of:
$$
(\alpha, \psi) \bar{\gamma} = (\alpha \circ p(\bar{\gamma}), R(\bar{\gamma}) \psi) \, , \quad \forall \bar{\gamma} \in \overline{G}\, , \psi \in \mathcal{H}_0 \, .
$$
It is a routine computation, similar to those performed in \cref{sec:inducing}, to check that the map $\pi_\mathcal{H} \colon \mathcal{H} \to \mathscr{M}$, given by $\pi_\mathcal{H}([\alpha, \psi]) = t(\alpha)$, is well defined, continuous, and defines a locally trivial bundle of Hilbert spaces. In fact, the map $i_\alpha \colon \mathcal{H}_0 \to \mathcal{H}$, $i_\alpha (\psi) = [\alpha, \psi]$, determines the appropriate identification of the fibres of $\pi_\mathcal{H}$ with the Hilbert space $\mathcal{H}_0$. The linear space structure on the fibres $\mathcal{H}_x = \pi_\mathcal{H}^{-1}(x)$ is induced by the identification provided by $i_\alpha$, that is:
\begin{equation}\label{eq:linearity}
[\alpha, \psi] + [\alpha, \phi] := [\alpha, \psi + \phi] \, , \qquad \lambda [\alpha, \psi] := [\alpha, \lambda \psi] \, ,
\end{equation}
for all $\psi, \phi \in \mathcal{H}_0$, $\alpha \colon x_0 \to x  \in \mathcal{G}$, $\lambda \in \mathbb{C}$;
and the inner product $\langle \cdot, \cdot \rangle_x$ on the fibres of $\pi_\mathcal{H}$ is defined in a natural way as:
$$ 
\langle [\alpha, \psi] , [\alpha, \phi] \rangle_x = \langle \psi, \phi \rangle_0 \, .
$$
It just remains to check that the projectivization of $\mathcal{H}$ is precisely the bundle $E_T$ previously constructed. This is trivial, because the ray corresponding to a vector $[\alpha, \psi]$ on the fibre $\mathcal{H}_x$ is defined as the ray $[[\alpha, \psi]] = \{\lambda [\alpha, \psi] \mid \lambda \neq 0 \}$, but because of \eqref{eq:linearity}, we get
$\lambda[\alpha, \psi] = [\alpha, \lambda\psi]$, and the ray $[[\alpha, \psi]] \in P(\mathcal{H}_x)$ agrees with the set $[\alpha, [\psi]] \in (E_T)_x$ (Eqs. (\ref{eq:projectivefibre}-\ref{eq:embedding})). Then we conclude that $P(\mathcal{H}) = E_T$.

\medskip

  The representation $\Phi$ preserves irreducibility since, if $\phi$ belongs to some invariant subspace $W \in \mathcal{H}_0$ of the representation $T$, then
\be
\Phi_\beta([\alpha,\, [\phi]]) \,=\, [\beta\circ\alpha,\, [\phi]] \,,
\ee
belongs to the subbundle
\be
W_T \,=\, \mathcal{G}_{x_0} \times P(W) /G_{x_0} \subset E_T \,.
\ee

Up to now, we have proved that any irreducible projective representation $\Phi$ of $\mathcal{G}$ restricts to an irreducible projective representation $\Phi_0$ of $G$, and that any irreducible projective representation $T$ of $G$ induces an irreducible projective representation of $\mathcal{G}$. It remains to prove that the restriction and the induction are compatible in the sense that:
\be \label{Eq: diagram compatibility}
r(\mathrm{Ind\,}(T)) = T \, , \qquad \Phi = \mathrm{Ind\, } (r(\Phi))
\ee
where $\Phi \in \mathrm{Proj\,}(\mathcal{G})$ and $T \in \mathrm{Proj}(G)$, with $\mathrm{Proj} (G)$ denoting the set of projective representations of $G$.

  We will show that \eqref{Eq: diagram compatibility} holds true by proving that the representation $\Phi'$ obtained out of $\Phi$ by first restricting to $G_{x_0}$ and then inducing on $\mathcal{G}$ is equivalent to $\Phi$ itself in the sense that there exists a map, say $\xi$, making the following diagram commutative, \cref{def:equivalence}:
\be \label{Eq: diagram equivalence}
\begin{tikzcd}
P(E_x) \arrow[d, "\Phi"]  & P(\mathcal{H}_x) \arrow[l, "\xi_x"] \arrow[d, "\Phi'"] \\
P(E_y) & P(\mathcal{H}_y) \arrow[l, "\xi_y"]
\end{tikzcd}
\ee
where $\mathcal{H} \to \mathscr{M}$ is the bundle supporting the representation $\Phi'$ and $E \to \mathscr{M}$ is the bundle supporting the representation $\Phi$.
The map
\be
\xi_x \;:\;\; P(\mathcal{H}_x) \to P(E_x) \;:\;\; [\alpha,\, [\phi]] \mapsto \xi_x([\alpha,\, [\phi]]) \,=\, \Phi_\alpha ([\phi]) \,,
\ee
makes \eqref{Eq: diagram equivalence} commutative as the following chains of equalities prove:
\begin{align}
\Phi_\beta \bigl(\,\xi_x ([\alpha,\, [\phi]])\,\bigr) \,&=\, \Phi_\beta \bigl(\,\Phi_\alpha ([\phi])\,\bigr) \,=\, \Phi_{\beta \circ \alpha} ([\phi]) \,, \\
\xi_y  \bigl(\Phi'_\beta([\alpha,\, [\phi]])\bigr)  \,&=\, \xi_y ([\beta \circ \alpha,\, [\phi]]) \,=\, \Phi_{\beta \circ \alpha} ([\phi]) \,.
\end{align}
The second identity in \eqref{Eq: diagram compatibility} is proved in a similar way.

To conclude the proof, we show that the previous constructions do not depend on the choice of the particular point $x_0$ fixed from the very beginning, namely, that the representation $\Phi'$ of $\mathcal{G}$ obtained out of a representation $\Phi$ by first restricting to $G_{x'_0}$ (for any other object $x'_0$ of $\mathcal{G}$) and then inducing on $\mathcal{G}$ is equivalent to the one obtained by first restricting to $G_{x_0}$ and then inducing on $\mathcal{G}$, in the sense that there exists a map $\xi$ such that the following diagram is commutative:
\be \label{Eq: diagram independence on x_0}
\begin{tikzcd}
P(\mathcal{H}_x) \arrow[r, "\xi_x"] \arrow[d, "\Phi"] & P(\mathcal{H}'_x) \arrow[d, "\Phi'"] \\
P(\mathcal{H}_y) \arrow[r, "\xi_y"] & P(\mathcal{H}'_y) 
\end{tikzcd}
\ee
where $\mathcal{H} \to \mathscr{M}$ is the bundle supporting $\Phi$ and $\mathcal{H}' \to \mathscr{M}$ is the bundle supporting $\Phi'$.
The map
\be
\xi_x \;:\;\; P(\mathcal{H}_x) \to P(\mathcal{H}'_x) \;:\;\; [\alpha,\, [\phi]] \mapsto \xi_x([\alpha,\, [\phi]]) \,=\, [\alpha \circ \sigma^{-1},\, [\phi]] \,,
\ee
for some (always existing, because of the transitivity of $\mathcal{G}$) morphism $\sigma$ connecting $x_0$ and $x'_0$, makes \eqref{Eq: diagram independence on x_0} commutative because of the following chains of equalities:
\begin{align}
\Phi'_\beta \bigl(\,\xi_x ([\alpha',\, [\phi]])\,\bigr) \,&=\, \Phi'_\beta ([\alpha' \circ \sigma^{-1},\,[\phi]]) \,=\, \\ 
\,&=\,[\beta \circ \alpha' \circ \sigma^{-1},\, [\phi]] \,=\, \nonumber \\
\,&=\, [\beta \circ \alpha,\, [\phi]] \,, \nonumber \\
\xi_y \Phi_\beta ([\alpha,\, [\phi]]) \,&=\, \xi_y [\beta \circ \alpha,\, [\phi]] \,=\, \\ 
\,&=\, [\beta \circ \alpha \circ \gamma,\, [\phi]] \,=\, \nonumber \\
\,&=\,[\beta \circ \alpha,\, [\phi]] \,, \nonumber 
\end{align}
for $\gamma \in G_{x_0}$.
\end{proof}

\bigskip

  Given a Lie groupoid $\mathcal{G}$ which is not necessarily transitive, because of \cref{Thm: Mackey theorem transitive groupoids}, each component $\Phi_O$ of the projective representation $\Phi$ (Eq. \eqref{eq:decomposition_rep}) is determined by the corresponding projective representations of the isotropy group $G_O$ of the orbit $O$.
An obvious corollary of \cref{Thm: Mackey theorem transitive groupoids}, using the definition of irreducible projective representation of a Lie groupoid given at the end of \cref{sec:projective} (\cref{rem:irreducible}), is:
\begin{corollary}[\textsc{Orbitwise Classification for Non-Transitive Lie Groupoids}] \label{Cor: Mackey theorem non-transitive groupoids}
Let \(\mathcal{G} \rightrightarrows \Omega\) be a Lie groupoid.
Then, the irreducible projective representations of \(\mathcal{G}\) are induced from irreducible projective representations $T_a \colon G_{O_a} \to PU(\mathcal{H}_0)$, where $\{ O_a \} = \Omega/\mathcal{G}$ is the collection of orbits of $\mathcal{G}$, and $T_a$ is an irreducible projective representation of the isotropy group \(G_{O_a}\) of the orbit $O_a$.
\end{corollary}


\section{Relativistic particles and relativistic groupoids: Wigner's classification of elementary particles revisited}
\label{Sec: Relativistic particles and relativistic groupoids: Wigner's elementary particles revisited}

In this section, we provide a new statement of \textit{Wigner's principle} that establishes a notion of an elementary particle for any Lorentzian space-time $(\mathscr{M},\,\eta)$.
Accordingly, we will apply \cref{Cor: Mackey theorem non-transitive groupoids} to the Wigner groupoid $\mathsf{Wigner}{(\mathscr{M},\,\eta)}$, obtaining a groupoidal based classification of elementary particles for generally curved Lorentzian space-times.

\begin{definition}[\textsc{Elementary particles in general space-times}] \label{Def: elementary particles in Lorentzian space-times}
Given a space-time $(\mathscr{M},\, \eta)$, elementary particles on it are described by irreducible projective representations of the connected component of its Wigner groupoid $\mathsf{Wigner}{(\mathscr{M},\, \eta)}$.
\end{definition}

 According to \cref{Cor: Mackey theorem non-transitive groupoids}, irreducible projective representations of $\mathsf{Wigner}{(\mathscr{M},\,\eta)}$ are classified by its orbits and by irreducible projective representations of the isotropy groups of each orbit.

  Orbits of $\mathsf{Wigner}{(\mathscr{M},\,\eta)}$ have already been discussed in \cref{Sec: Relativistic groupoids: The Wigner groupoid}, \cref{prop:foliation} (see Fig. \ref{fig:orbits}).
They are parametrized by the square norm of the covector $p \in  \mathbf{T}^* \mathscr{M}$ and they can be collected into two main families: $\mathcal{O}_{m,+}$, those for which $p^2 \,>\, 0$ ($p^2 \,=:\,m^2$ for $m \in \mathbb{R}_+$), and $\mathcal{O}_{0,+}$, those for which $p^2 \,=\, 0$ and $p_0 \,>\, 0$. 

\subsection{Massive particles}
A representative of the orbit $\mathcal{O}_{m,+}$ selected by $p^2 \,=\, m^2$ can be taken to be:
$$
\zeta_0 = (0, ... ,\,0;\,m,\,0, ... ,\,0,\,) \in  \mathbf{T}^* \mathscr{M} \,.
$$
Its isotropy group is the space of morphisms of the connected component of $\mathsf{Wigner}{(\mathscr{M},\,\eta)}$ preserving the point $\zeta_0$, which amounts to linear transformations from $\mathbf{T}_{x_0} \mathscr{M}$ to $\mathbf{T}_{x_0} \mathscr{M}$ belonging to $SO(n-1)$.
Thus, by virtue of \cref{Thm: Mackey theorem transitive groupoids}, irreducible projective representations of $\mathsf{Wigner}{(\mathscr{M},\, \eta)}$ are labelled, for positive values of the parameter $m$, by irreducible projective representations of $SO(n-1)$. In the particular instance of $n = 4$, we get that irreducible projective representations of $SO(3)$ are determined by irreducible unitary representations of the universal covering of $SO(3)$, that is, $SU(2)$, as in Wigner's classification.

\subsection{Massless particles}\label{sec:massless}
On the other hand, the orbit $\mathcal{O}_{0,+}$ with $p^2 \,=\, 0$ and $p_0 \,>\, 0$, can be represented by the point:
$$
\zeta_1 \,=\, (0,...,\,0;\,E,\,E,\,0,...,0) \in  \mathbf{T}^* \mathscr{M} \, .
$$
Its isotropy group is the space of morphisms preserving such a point, which amounts to linear transformations from $\mathbf{T}_{x_0} \mathscr{M}$ to $\mathbf{T}_{x_0} \mathscr{M}$ belonging to $ISO(n-2)$.
Thus, for the value $0$ of the parameter $m$ labelling the orbits, irreducible projective representations of $\mathsf{Wigner}{(\mathscr{M},\, \eta)}$ are labelled by irreducible projective representations of the inhomogeneous orthogonal group $ISO(n-2)$, also known as the Euclidean group in dimension $n-2$ and denoted as $E(n-2)$.

We will concentrate on the four-dimensional scenario, which can be extended to higher dimensions.
In such a case, the isotropy group of the massless particle orbit $\mathcal{O}_{0,+}$ is the Euclidean group in 2 dimensions, that is $E(2) := ISO(2)$, which is a connected three-dimensional Lie group. Hence, massless particles are classified according to the irreducible projective representations of $E(2)$. In what follows, we study the projective unitary representations of the Euclidean group
\(
E(2)=\mathbb{R}^2 \rtimes SO(2).
\)
To do that, we will follow the approach of Cari\~nena--Santander \cite{Carinena-Santander-ProjectiveUnitary-1975} stated before, \cref{sec:projective}, \cref{fig:extension}: Every projective representation of $E(2)$ is associated with a unitary representation of its \emph{projective covering group} $\overline{E(2)}$. It will be shown in what follows that $\overline{E(2)} = H(2) \rtimes \mathbb{R}_\phi$, with $H(2)$ the Heisenberg group. The classification is then obtained by applying Mackey's general theory of induced representations for semidirect products, with $H(2)$ as the normal subgroup.

\subsubsection{The projective covering group $\overline{E(2)}$}

Following the prescription in \cite{Carinena-Santander-ProjectiveUnitary-1975}, we first compute the second cohomology group $H^2(\mathfrak{e}(2))$ of the Lie algebra of $E(2)$. The Lie algebra $\mathfrak e(2)$ has generators $J,P_1,P_2$ with brackets:
\begin{equation*}
[J,P_1]=P_2,\qquad [J,P_2]=-P_1,\qquad [P_1,P_2]=0.
\end{equation*}

  A direct calculation of the Chevalley--Eilenberg cohomology gives\footnote{Every skew-symmetric $\omega \colon \mathfrak e(2) \times \mathfrak e(2) \to \mathbb{R}$ satisfies $\delta \omega = 0$, so it is a 2-cocycle. Then, if $\alpha \colon \mathfrak e(2) \to \mathbb{R}$, $\delta \alpha = a P_1\wedge J + b P_2 \wedge J$, then $Z^2/B^2 = \{ \mu P_1 \wedge P_2 \mid \mu \in \mathbb{R} \}$ (we are using the same symbols $P_1$, $P_2$, $J$ for the linear basis of $\mathfrak e(2)$ and its dual basis).}:
\begin{equation*}
H^2(\mathfrak e(2),\mathbb R) \cong \mathbb R,
\end{equation*}
with the generator $\omega$ defined by
\begin{equation*}
\omega(P_1,P_2)=1, \qquad \omega(J,P_i)=0.
\end{equation*}

  Hence, the projective covering Lie algebra is \cite{Carinena-Santander-ProjectiveUnitary-1975}:
\begin{equation*}
\overline{\mathfrak e(2)} = \mathbb{R}^{\dim H^2(\mathfrak e(2),\mathbb R) } \oplus  \mathfrak e(2) = \mathbb R E \oplus \mathfrak e(2),
\end{equation*}
with commutators
\begin{equation*}
[E,\cdot]=0,\quad [P_1,P_2]=E,\quad [J,P_i]=\varepsilon_{ij}P_j.
\end{equation*}

  Its connected, simply connected Lie group $\overline{E(2)}$ is the projective covering of the Euclidean group $E(2)$.

  The Lie algebra of $\overline{E(2)}$ is isomorphic to the Lie algebra of the oscillator group \cite{St67}; hence $\overline{E(2)}$ is its universal covering. The oscillator group is a 4-dimensional connected solvable Lie group, hence the same applies to $\overline{E(2)}$.

  Using the explicit parametrization of the oscillator group in \cite{St67}, we get
a global parametrization of $\overline{E(2)}$ given by:
\begin{equation*}
(s,a,\phi),\qquad s\in\mathbb R,\ a=(a_x,a_y)\in\mathbb R^2,\ \phi\in\mathbb R,
\end{equation*}
with multiplication law
\begin{equation}\label{eq:product}
(s,a,\phi)(s',a',\phi')
=\Big(s+s'+\tfrac{1}{2}\,a\times R_\phi a',\ a+R_\phi a',\ \phi+\phi'\Big),
\end{equation}
where $a\times b = a_x b_y - a_y b_x$ and $R_\phi$ is a rotation by angle $\phi$. The group product \eqref{eq:product} can be derived from the Baker-Campbell-Hausdorff formula using the fact that $[P_1,P_2]=E$ is central, producing the $\tfrac12\,a\times R_\phi a'$ term.

  The canonical projection homomorphism $p \colon \overline{E(2)} \to E(2)$ is given by:
$$
p(s,a,\phi) = (a, R_\phi ) \, , \qquad R_\phi = \left(\begin{array}{rr} \cos \phi & - \sin \phi \\ \sin \phi & \cos \phi \end{array} \right) \, ,
$$
and $\ker p = \{ (s, 0, \pi k) \mid s \in \mathbb{R}, k \in \mathbb{Z} \}$.

  Note that the elements of the group of the form $(s,a,0)$ form a normal subgroup isomorphic to the Heisenberg group $H(2)$. It is also easy to check that $H(2)$ is a normal subgroup of $\overline{E(2)}$. Finally, the elements of the form $(0,0,\phi)$ form a subgroup $\mathbb{R}_\phi$ of $\overline{E(2)}$ acting naturally on $H(2)$ by $\phi \cdot (s,a) = (s, R_\phi a)$. Then $\overline{E(2)}$ is the semidirect product:
\begin{equation*}
\overline{E(2)} = H(2)\rtimes \mathbb R_\phi,
\end{equation*}
where $H(2)$ is the three-dimensional Heisenberg group generated by $(E,P_1,P_2)$ and $\mathbb R_\phi$ is the universal cover of $SO(2)$.

  We also write $\overline{E(2)}=N\rtimes H$ with $N=H(2)$ and $H=\mathbb R_\phi$.


\subsubsection{Mackey's Theory for semidirect products}

We would like to compute now the irreducible unitary representations $R$ of the projective covering $\overline{E(2)}$ and, among them, select those such that $R(\ker p) \subset U(1)$. To do that, we will take advantage of the fact that $\overline{E(2)} = N \rtimes H$ is a semidirect product, so we can use Mackey's theory. However, we cannot apply the simple version of Mackey's machine valid when $N$ is an Abelian group, Thm. \ref{Thm: Mackey semidirect}, because now $N = H(2)$ is not Abelian. However, there is a particular instance of general theorems by Mackey on the theory of representations of extensions of groups \cite{mackey1958unitary} that applies nicely to this situation. We will write down explicitly the theorem needed in the present context.

\begin{theorem}\label{thm:non_abelian_extension}
Let $G = N \rtimes H$ be a connected locally compact topological group which is the semidirect product of $N$, a closed type I normal subgroup, and $H = G/N$. Then, all irreducible unitary representations $R \in \widehat{G}$ are obtained from pairs $(\pi, \sigma)$, where $\pi\in\widehat N$ is an irreducible unitary representation of $N$, and $\sigma \in \widehat{H}_\pi$ is an irreducible unitary representation of the isotropy group $H_\pi$ corresponding to the canonical action of $H$ on $\widehat{N}$. The representation $R$ is the induced representation of the representation $\tilde{\pi} \otimes \sigma$, that is: $R = \mathrm{Ind}_{N\otimes H_\pi}^G (\tilde{\pi}\otimes \sigma)$, with $\tilde{\pi}$ an extension of the representation $\pi$ to $N \rtimes H_\chi$ such that $\tilde{\pi}\mid_N = \pi$. Moreover, if $\pi$, $\pi'$ are in the same orbit of $H$ and $\sigma$, $\sigma'$ are equivalent representations, then the corresponding induced representations $R$ and $R'$ are equivalent.
\end{theorem}

Thus, the first step to apply Mackey's theorem to the projective covering group $\overline{E(2)}$ is to analyze the space $\widehat{H(2)}$ of irreducible unitary representations of the Heisenberg group $H(2)$ and the orbits of $\mathbb{R}_\phi$ on it.


\subsubsection{Dual of $H(2)$}

The computation of irreducible unitary representations of the Heisenberg group is done more easily using Kirillov's theory. Indeed, $H(2)$ is a nilpotent Lie group of dimension three. Then, Kirillov's theorem states that there is a one-to-one correspondence between $\widehat{H(2)}$ and the space of coadjoint orbits of its Lie algebra $\mathfrak{h}(2)$. The corresponding unitary representations are obtained by quantizing such coadjoint orbits.
After some computations, using the dual of the adjoint action on the generators $E, P_1, P_2$ of $\mathfrak{h}(2)$, we get that there is a family of two-dimensional coadjoint orbits parametrized by a real number $\mu \neq 0$ (the value of the Casimir $E$), while for $\mu = 0$, the orbits are zero-dimensional, i.e., just points $(p_1, p_2) \in \mathfrak{h}(2)^*$. Thus, the unitary dual $\widehat{H(2)}$ has two strata:
\begin{enumerate}
\item Central character $\mu=0$: coadjoint orbits are points $p = (p_1,p_2)$ and the corresponding unitary representations are the one-dimensional characters:
$$
\chi_p(s,a)=e^{i\,p\cdot a} = e^{i(a_xp_x + a_yp_y)}\, , \quad p\in\mathbb R^2.
$$
\item Central character $\mu\ne 0$: coadjoint orbits are 2-dimensional and their quantization, by the Stone--von Neumann theorem, is provided by the unique (up to equivalence) irreducible representation $\pi_\mu$ on $L^2(\mathbb R)$ given by:
\begin{equation}\label{eq:pimu}
\pi_\mu(s,0,0)=e^{i\mu s},\quad \pi_\mu (0,a_x,0) = e^{i a_x Q} \, , \quad \pi_\mu (0,0,a_y) = e^{ia_yP}  
\end{equation}
with $Q, P$ the standard position and momentum operators on $L^2(\mathbb{R})$, $Q \psi (u) = u \psi (u)$, $P = - i \mu \frac{d}{du}$.
\end{enumerate}
Note that in this representation the generators $P_1, P_2$ of the Lie algebra $\mathfrak{h}(2)$ are represented by operators $\hat{P}_1, \hat{P}_2$, such that:
$$
[\hat{P}_1,\hat{P}_2]=i\mu\,\mathbf 1 \,  .
$$
The constant $\mu$ must be understood as a ``magnetic field'', not as a ``$\hbar$''-like constant, because the generators $P_1$, $P_2$ are the momenta corresponding to translations on the $x$ and $y$ directions on the space of momenta $\mathfrak{h}(2)^*$  respectively. Thus, even if in the representation $\pi_\mu$ they appear as ``position'' and ``momentum'' operators, Eq. (\ref{eq:pimu}), physically they are both momenta in two dimensions, so the representations with $\mu \neq 0$ could be called a ``magnetic'' sector.

Clearly, the action of $\mathbb{R}_\phi$ on $\widehat{H(2)}$ is given by:
$$
\phi\cdot \chi_p = \chi_{R_\phi p}\, ,
$$
for representations $\chi_p$ corresponding to $\mu = 0$; that is, they draw circles of radius $\rho = || p ||$ in the plane $\mu = 0$. If $p \neq 0$, then the isotropy group $H_{\chi_p}$ is trivial. If $p = 0$, then the isotropy group is $H = \mathbb{R}_\phi$ itself.

  On the other hand, if $\mu \neq 0$, then:
$$
\phi\cdot \pi_\mu = \pi_\mu.
$$
and the isotropy group is $H = \mathbb{R}_\phi$.

  Thus:
\begin{enumerate}
\item For $\mu=0$, the orbits are $\{0\}$ (whose isotropy group is $\mathbb{R}_\phi$),
and the circles $|p|=\rho>0$  (whose isotropy groups are given by integer multiples of $2\pi$).

\item For $\mu\ne 0$, each point $\pi_\mu$ is a fixed orbit  (whose isotropy group is $\mathbb{R}_\phi$).
\end{enumerate}

  Now, we can apply Mackey's theorem by taking the products $\tilde{\pi} \otimes \sigma$, where $\pi \in \widehat{H(2)}$ and $\sigma \in \widehat{\mathbb{R}}_\phi$. Because $\mathbb{R}_\phi$ is Abelian, $\sigma$ will be a character $\chi_\lambda(\phi) = e^{i\lambda\phi}$, $\lambda\in \mathbb{R}$. On the other hand, $\pi$ will fall into one of the two classes of representations discussed above, $\mu = 0$ and $\mu \neq 0$. In the magnetic sector $\mu \neq 0$, the isotropy group is the total group $H$, so there is no need for induction and $\tilde{\pi}_\mu \otimes \chi_\lambda$ is the desired representation. However, in the sector $\mu = 0$, there are two situations with different isotropy groups.

\subsubsection{Sector $\mu=0$ (ordinary $E(2)$ representations)}

As we mentioned above, if $\mu = 0$, there are two possibilities:

\begin{enumerate}
\item $p=0$: the stabilizer $H_{\chi_0}$ is all of $H=\mathbb R_\phi$, so again there is no need for induction and the irreducible unitary representations are given by $\pi_0 \otimes \chi_\lambda$, that is, by the one-dimensional representations:
$$
(s,a,\phi)\mapsto \pi_0 \otimes \chi_\lambda (s,a,\phi) = e^{i0.a}e^{i\lambda\phi} = e^{i\lambda\phi} \, .
$$
Now, these representations all produce representations of the universal covering $\widetilde{E(2)}$, but to provide projective representations of $E(2)$ they must descend to $SO(2)$ on the factor $\phi$, which implies that $\lambda \in \mathbb{Z}$. Thus, the representations in the sector $\mu = 0$, with $p = 0$, provide the projective representations of $E(2)$ with integer helicity $\lambda  \in \mathbb{Z}$ in Wigner's standard classification.

\item $|p|=\rho>0$: orbit is the circle $S^1_\rho$. The isotropy group (or stabilizer group or even ``little group'') is $2\pi\mathbb Z$, $H_{\chi_p} = 2\pi \mathbb{Z}$. Then the dual group of $\mathbb{Z}$ is $U(1)$; that is, for each $e^{i\varphi} \in U(1)$, we have the representation $2\pi k \mapsto \sigma_\varphi (2\pi k) = e^{2\pi i k \varphi}$. Characters  $\sigma$ of the little group are thus labelled $\sigma_\varphi$, $\varphi \in [0,2\pi)$. The induced representation corresponding to $\pi_{0,p} \otimes \sigma_\varphi$ acts on $L^2(S^1)$ (note that $\overline{E(2)} / H(2) \rtimes 2\pi \mathbb{Z} = U(1) \cong S^1$), by:
\begin{align*}
(T_a f)(\theta)&=e^{i\rho\,a\cdot \hat n(\theta)}f(\theta),\qquad \hat n(\theta)=(\cos\theta,\sin\theta),\\
(R_\phi f)(\theta)&=e^{i\varphi\phi}f(\theta-\phi).
\end{align*}

All these representations pass to the quotient and they provide genuine projective representations of $E(2)$. They correspond to the continuous spin representations in Wigner's standard classification \cite{Gracia-Bondia-2018}.
\end{enumerate}

Thus, we conclude that the sector $\mu = 0$ reproduces the standard classification, except for half-integer helicity massless particles, in Wigner's standard approach.

\subsubsection{Sector $\mu\ne 0$ (magnetic representations)}

Consider $\pi_\mu$ given by the standard Stone--von Neumann representation on $L^2(\mathbb{R})$. The stabilizer is $H=\mathbb R_\phi$, so no induction is needed, but $\pi_\mu$ must be extended to a representation $\tilde{\pi}_\mu$ on the whole $H(2) \rtimes \mathbb{R}_\phi$. This is done via the standard representation of the harmonic oscillator (a representation of the oscillator group \cite{St67}).

\begin{theorem}
For each $\mu\ne 0$, there exists an irreducible representation $\tilde{\pi}_\mu$ of $\overline{E(2)}$ on $L^2(\mathbb R)$ with:
\begin{equation*}
\hat{E} =\mu\,\mathbf 1,\quad
\hat{P}_x=\sqrt{|\mu|}\,Q,\quad \hat{P}_y=\operatorname{sgn}(\mu)\sqrt{|\mu|}\,P,\quad
\hat{J}=\tfrac12(Q^2+P^2),
\end{equation*}
where $Q$ is the position operator (multiplication by $x$), and $P=-i\,\frac{d}{dx}$.
\end{theorem}

  The operator $\hat{J}$ has spectrum $\{n+\tfrac12: n\in\mathbb N_0\}$, the harmonic oscillator spectrum. All the remaining representations obtained by using non-trivial characters of $\mathbb{R}_\phi$, $\chi_{c_0} (\phi) = e^{ic_0 \phi}$, correspond to a shift in the zero ``energy'' of the harmonic oscillator, that is, shifting the spectrum of $\hat{J}$ by $c_0$.

\bigskip

Thus, we can conclude that:

\begin{proposition}\label{prop:classification_E(2)}
Irreducible projective representations of $E(2)$ are classified according to the following list.

\medskip

  Ordinary sector ($\mu=0$):
\begin{enumerate}
\item (Integer valued helicities) One-dimensional characters $e^{i\lambda\phi}$, $\lambda \in \mathbb{Z}$.
\item (Continuous spin representations) Induced circle representations on $L^2(S^1)$ with parameters $\rho>0$ and $\varphi\in\mathbb R/\mathbb Z$.
\end{enumerate}

  Magnetic sector ($\mu\ne 0$):
\begin{enumerate}
\item For each $\mu\in\mathbb R\setminus\{0\}$, a family of irreducible representations $\Pi_{\mu, c_0}$ on $L^2(\mathbb R)$ (or $\mathcal H_B$) with $[\hat{P}_1,\hat{P}_2]=i\mu\,\mathbf 1$ and $\sigma(\hat{J})=N+\tfrac12 + c_0$.
\end{enumerate}
\end{proposition}

\begin{remark}
The previous classification holds for any space-time $\mathscr{M}$; thus, the fundamental structural properties of elementary particles do not depend on the space-time being a homogeneous space of the Poincar\'e group or any other relativistic group. The structural properties of elementary particles depend solely on the causal structure of the space-time.
\end{remark}

\begin{remark}
The obtained classification recovers Wigner's classification of half-integer spin massive particles, integer helicity massless particles, and continuous spin representations and the previous remark explaining the robustness of the classification of elementary particles provided by the quantum numbers $(m,s)$.
\end{remark}

\begin{remark}
What is new with respect to Wigner's classification is that there is a new family of massless particles characterized by a magnetic moment-like $\mu \neq 0$. It remains to determine whether or not this magnetic sector represents a new genuine family of particles or if it is just a mathematical artifact that should be ruled out for physical reasons.   

Note the the sector $\mu \neq 0$ does not appear in standard classification of elementary particles because when using the Poincar\'e group as the global symmetry to characterize elementary particles, its irreducible projective representations are in one-to-one correspondence with the irreducible unitary representations of its double covering.  Then, when applying Mackey's machine, massless particles are in one-to-one  correspondence with irreducible \textbf{unitary} representations of the double covering $\widetilde{E(2)}$ of the Euclidean group $E(2)$, the isotropy group of the momenta $p$ with $p^2 = 0$.  However when we start from the Wigner groupoid of the given space-time massless particles are in one-to-one correspondence with irreducible \textbf{projective} representations of the Euclidean group $E(2)$, the isotropy group of the orbit $\mathcal{O}_{0,+}$, and this family is richer that just unitary representations.   Indeed irreducible projective representations of $E(2)$ are in one-to-one correspondence with irreducible unitary representations of central extension $\overline{E(2)}$ of $E(2)$ which is characterized by the parameter $\mu$.   When $\mu = 0$ we recover the well-known continuous spin representations, however if $\mu \neq 0$, we obtain a genuinely new family of representations describing new massless particles.
\end{remark}

\begin{remark}
If $(\mathscr{M},\eta)$ admits a spin structure, one can replace the orthonormal
frame bundle by its spin double covering and thus lift the Wigner groupoid
to a $\mathrm{Spin}(1,3)$-groupoid. In this way, irreducible projective
representations of $\mathrm{Spin}(1,3)$ yield fermionic particles on the massless orbit, and
the existence of such representations is controlled by the vanishing of
the second Stiefel-Whitney class $w_2(\mathscr{M})$. More generally, if $\mathscr{M}$ only
admits a $\mathrm{Spin}^c$ structure, one can incorporate an additional
$U(1)$ factor.  We leave the discussion of these issues to a forthcoming publication.
\end{remark}


\section{Conclusions, discussion and outlook}\label{sec:conclusions}

In this work, we have extended Wigner's program to the generic situation of space-times which are not homogeneous spaces of kinematical groups. In this way, we have provided a natural background for the notion of elementary particles in curved space-times. In particular, a Wigner-like classification of elementary particles is obtained for a large class of space-times $\mathscr{M}$.  

These results are obtained by replacing the group of isometries of the given space-time $\mathscr{M}$ with its Wigner groupoid $\mathsf{Wigner}(\mathscr{M}, \eta)$, a much more flexible geometrical structure than the group of isometries, which provides a natural framework for the notion of ``symmetries'' of $\mathscr{M}$. Then, following Wigner's program, the notion of an elementary particle on $\mathscr{M}$ is identified with the irreducible projective representations of $\mathsf{Wigner}(\mathscr{M}, \eta)$.

A first step towards the goal of classifying elementary particles for arbitrary space-times has been taken by establishing an extension of the theory of induction of projective representations for Lie groups to Lie groupoids. It has been shown that there is a one-to-one correspondence between irreducible projective representations of the connected components of Lie groupoids and irreducible projective representations of their isotropy groups.

As a consequence of all this, for a given space-time $\mathscr{M}$, we recover Wigner's standard classification of elementary particles, except for the fact that half-integer helicity massless particles are not present and a new family of representations arises, which is parametrized by an intrinsic magnetic-like moment.  This result stands up even for Minkowski space-time.   A similar analysis can be conducted for spacetimes in dimension different than four and other symmetry groups.   These aspects will be discussed in subsequent works.

It is relevant to stress here that the proof of the main theorem, \cref{Thm: Mackey theorem transitive groupoids}, presented here does not require restrictive regularity hypotheses on the given groupoids and can be extended to a larger class of groupoids, including diffeological groupoids. The details of this will be discussed elsewhere.

A finer analysis of the theory of projective representations of Lie groupoids and its relation with the theory of extensions of groupoids \cite{Renault-GroupoidApproach-1980} will be addressed elsewhere, including a characterization of projective representations of groupoids in terms of unitary representations of a universal covering groupoid in the spirit of \cite{Carinena-Santander-ProjectiveUnitary-1975}.

The implications of the present work regarding long-standing problems in quantum field theories, like the high-spin problem (an instance of a family of no-go theorems, like the Coleman-Mandula theorem, that restrict the class of symmetries or particles that can be described in relativistic quantum field theories), are not addressed in this work. However, because of their deep significance, they deserve to be analyzed from the perspective offered by the introduction of this new notion of symmetry.

Finally, it must be pointed out that the Poincar\'e groupoid would allow us to address the problem of extending the notion of covariant representations and the corresponding notion of relativistic covariance for field theories on curved space-times. We expect that they will be helpful in understanding the problem of constructing relativistic equations for higher spin particles.


\section*{Acknowledgements and funding}
A.I. and A.M. acknowledge financial support from the Spanish Ministry of
Economy and Competitiveness, through the Severo Ochoa Program for Centers of Excellence in RD
(SEV-2015/0554), the MINECO research project PID2024-160539NB-I00, and the Comunidad de
Madrid project TEC-2024/COM-84 QUITEMAD-CM.  

  L.S. acknowledges financial support from Next Generation EU through the project 2022XZSAFN PRIN2022 CUP: E53D23005970006.
L.S. is a member of the GNSGA (Indam).

\section*{Conflict of interest} The authors have no conflicts of interest to disclose.


\begin{thebibliography}{99}

\bibitem[Ba54]{Bargmann-UnitaryRay-1954}
V.~Bargmann.
 {U}nitary {R}ay {R}epresentations of {C}ontinuous {G}roups.
 {\em Annals of Mathematics}, \textbf{59} (1), 1--46, 1954.

\bibitem[Bl13]{Blohmann2013}
C.~Blohmann, M.~Fernandes, and A.~Weinstein.
 Groupoid symmetry and constraints in general relativity.
 {\em Communications in Contemporary Mathematics}, \textbf{15} (01), January
  2013.

\bibitem[Ba15]{Bautista-Ibort-Causality-2015}
A.~Bautista, A.~Ibort, and J.~Lafuente.
 Causality and skies: is non-refocussing necessary?
 {\em Classical and Quantum Gravity}, \textbf{32}, 105002, 2015.

\bibitem[Ba68]{Bacry-LevyLeblond-Kinematics-1968}
H.~Bacry and J.~M. L{\'e}vy-Leblond.
 Possible kinematics.
 {\em Journal of Mathematical Physics}, \textbf{9} (10), 1605--1614, 1968.

\bibitem[Bo06]{Bos-ContinuousRepresentations-2006}
R.~Bos.
 {C}ontinuous representations of groupoids.
 {\em arXiv:math / 0612639 [math.RT]}, 2006.

\bibitem[Bo11]{bos2011continuous}
R.~Bos.
 Continuous representations of groupoids.
 {\em Houston journal of mathematics}, 37(3):807--844, 2011.

\bibitem[Ci21]{Ciaglia-DiCosmo-Ibort-Marmo-Schiavone-Symmetroids-2021}
F.~M. Ciaglia, F.~Di~Cosmo, A.~Ibort, G.~Marmo, and L.~Schiavone.
 {S}chwinger's picture of {Q}uantum {M}echanics: 2-groupoids and
  symmetries.
 {\em Journal of Geometric Mechanics}, 13(3):333--357, 2021.

\bibitem[Ci21b]{Ciaglia+-DynamicalMaps-2021}
F.~M. Ciaglia, F.~Di~Cosmo, A.~Ibort, and G.~Marmo.
 {D}ynamical maps and {S}ymmetroids.
 {\em Open Systems and Information Dynamics}, 28(04):2150019, 2021.

\bibitem[Ca75]{Carinena-Santander-ProjectiveUnitary-1975}
J.~F. Cari{\~n}ena and M.~Santander.
 {O}n the projective unitary representations of connected {L}ie
  groups.
 {\em Journal of Mathematical Physics}, 16(7):1416--1420, 1975.

\bibitem[Di63]{Dixmier-Douady-ChampsContinus-1963}
J.~Dixmier and A.~Douady.
 {C}hamps continus d'espaces hilbertiens et de {C}$^*$-alg{\`e}bres.
 {\em Bulletin de la Soci{\'e}t{\'e} Math{\'e}matique de France},
  91:227--284, 1963.

\bibitem[Fo16]{Folland-HarmonicAnalysis-2016}
G.~B. Folland.
 {\em A course in abstract harmonic analysis}.
 CRC press, Boca Raton, FL, 2016.

\bibitem[Gr18]{Gracia-Bondia-2018}
J.~M. Gracia-Bondia, F.~Lizzi, J.~C. Varilly, and P.~Vitale.
 The kirillov picture for the wigner particle.
 {\em Journal of Physics A: Mathematical and Theoretical}, 51:255203,
  2018.

\bibitem[He05]{Heller-Pys-Sasin-NoncommutativeUnification-2005}
M.~Heller, L.~Pysiak, and W.~Sasin.
 Noncommutative unification of general relativity and quantum
  mechanics.
 {\em Journal of Mathematical Physics}, 46:122501--15, 2005.

\bibitem[He07]{Heller-Pys-Sasin-ConceptualUnification-2007}
M.~Heller, L.~Pysiak, and W.~Sasin.
 Conceptual unification of gravity and quanta.
 {\em International Journal of Theoretical Physics}, 46:2494--2512,
  2007.

\bibitem[Ib19]{Ibort-Rodriguez-IntroGroupoids-2019}
A.~Ibort and M.~A. Rodr{\'\i}guez.
 {\em An {I}ntroduction to {G}roups, {G}roupoids and {T}heir
  {R}epresentations}.
 CRC Press, Boca Raton, 2019.

\bibitem[Lo89]{Low-NullGeodesics-1989}
R.~J. Low.
 The geometry of the space of null geodesics.
 {\em Journal of Mathematical Physics}, 30:809--811, 1989.

\bibitem[Ma52]{mackey1952}
G.W. Mackey.
 Induced representations of locally compact groups i.
 {\em Ann. of Maths.}, 55(1):101--139, 1952.

\bibitem[Ma58]{mackey1958unitary}
G.W. Mackey.
 Unitary representations of group extensions i.
 {\em Acta Mathematica}, 99(1):265--311, 1958.

\bibitem[Ma05]{Mackenzie-LieGroupoids-2005}
K.~C.~H. Mackenzie.
 {\em {G}eneral {T}heory of {L}ie {G}roupoids and {L}ie {A}lgebroids}.
 Cambridge University Press, Cambridge, 2005.

\bibitem[Py04]{Pysiak-ImprimitivityGroupoids-2004}
L.~Pysiak.
 Groupoids, their representations and imprimitivity systems.
 {\em Demonstratio Mathematica}, 37:661--670, 2004.

\bibitem[Py11]{Pysiak-ImprimitivityGroupoids-2011}
L.~Pysiak.
 Imprimitivity theorems for groupoid representations.
 {\em Demonstratio Mathematica}, 44:29--48, 2011.

\bibitem[Re80]{Renault-GroupoidApproach-1980}
J.~Renault.
 {\em {A} groupoid approach to {C}$^*$-algebras}, volume 793 of {\em
  Lecture Notes in Mathematics}.
 Springer-Verlag, Berlin, Heidelberg, 1980.

\bibitem[Si68]{simms1968lie}
D.~J. Simms.
 {\em Lie groups and quantum mechanics}, volume~52.
 Springer-Verlag, Berlin, Heidelberg, 1968.

\bibitem[St67]{St67}
R.F. Streater.
 The representations of the oscillator group.
 {\em Commun. Math. Phys.}, \textbf{4}, 217--236, 1967.

\bibitem[St00]{streater2000pct}
R.~F. Streater and A.~S. Wightman.
 {\em PCT, spin and statistics, and all that}, volume~30.
 Princeton University Press, 2000.

\bibitem[Wa84]{Wald-GeneralRelativity-1984}
R.~M. Wald.
 {\em {G}eneral {R}elativity}.
 University of Chicago Press, Chicago, 1984.

\bibitem[We95]{Weinberg-QFT1-1995}
S.~Weinberg.
 {\em {T}he {Q}uantum {T}heory of {F}ields 1: {F}oundations}.
 Cambridge University Press, Cambridge, 1995.

\bibitem[We96]{Weinberg-QFT2-1996}
S.~Weinberg.
 {\em {T}he {Q}uantum {T}heory of {F}ields 2: {M}odern
  {A}pplications}.
 Cambridge University Press, Cambridge, 1996.

\bibitem[We96b]{Weinstein-GroupoidsSymmetry-1996}
A.~Weinstein.
 Groupoids: {U}nifying {I}nternal and {E}xternal {S}ymmetry.
 {\em Notices of the American Mathematical Society}, \textbf{43} (7), 744--752,
  1996.

\bibitem[Wi39]{Wigner-UnitaryRepresentations-1939}
E.~P. Wigner.
 {O}n {U}nitary {R}epresentations of the {I}nhomogeneous {L}orentz
  {G}roup.
 {\em Annals of Mathematics}, \textbf{40} (1), 149--204, 1939.

\bibitem[Wi22]{witten2022does}
E.~Witten.
 Why does quantum field theory in curved spacetime make sense? and
  what happens to the algebra of observables in the thermodynamic limit?
 In {\em Dialogues Between Physics and Mathematics: CN Yang at 100},
  pages 241--284. Springer, 2022.
\end{thebibliography}

\newcommand{\etalchar}[1]{$^{#1}$}

\end{document}